\documentclass[12pt,english,sort&compress]{elsarticle}
\usepackage[T1]{fontenc}
\usepackage[latin9]{inputenc}
\usepackage[letterpaper]{geometry}
\usepackage{amsbsy}
\usepackage{amstext}
\usepackage{graphicx}
\usepackage{amssymb}
\usepackage{esint}

\makeatletter

\providecommand{\tabularnewline}{\\}
\newcommand{\lyxdot}{.}

\usepackage{babel}

\begin{document}

\title{An Energy- and Charge-conserving, Implicit, Electrostatic Particle-in-Cell Algorithm}

\author{G. Chen, L. Chacón, }

\address{Oak Ridge National Laboratory, Oak Ridge,TN 37831, USA }

\author{D. C. Barnes}

\address{Coronado Consulting, Lamy, NM 87540, USA }
\begin{abstract}
This paper discusses a novel fully implicit formulation for a one-dimensional
electrostatic particle-in-cell (PIC) plasma simulation approach. Unlike
earlier implicit electrostatic PIC approaches (which are based on
a linearized Vlasov-Poisson formulation), ours is based on a nonlinearly
converged Vlasov-Ampère (VA) model. By iterating particles and fields
to a tight nonlinear convergence tolerance, the approach features
superior stability and accuracy properties, avoiding most of the accuracy
pitfalls in earlier implicit PIC implementations. In particular, the
formulation is stable against temporal (Courant-Friedrichs-Lewy) and
spatial (aliasing) instabilities. It is charge- and energy-conserving
to numerical round-off for arbitrary implicit time steps (unlike the
earlier {}``energy-conserving'' explicit PIC formulation, which
only conserves energy in the limit of arbitrarily small time steps).
While momentum is not exactly conserved, errors are kept small by
an adaptive particle sub-stepping orbit integrator, which is instrumental
to prevent particle tunneling (a deleterious effect for long-term
accuracy). The VA model is orbit-averaged along particle orbits to
enforce an energy conservation theorem with particle sub-stepping.
As a result, very large time steps, constrained only by the dynamical
time scale of interest, are possible without accuracy loss. Algorithmically,
the approach features a Jacobian-free Newton-Krylov solver. A main
development in this study is the nonlinear elimination of the new-time
particle variables (positions and velocities). Such nonlinear elimination,
which we term particle enslavement, results in a nonlinear formulation
with memory requirements comparable to those of a fluid computation,
and affords us substantial freedom in regards to the particle orbit
integrator. Numerical examples are presented that demonstrate the
advertised properties of the scheme. In particular, long-time ion
acoustic wave simulations show that numerical accuracy does not degrade
even with very large implicit time steps, and that significant CPU
gains are possible. 
\end{abstract}
\maketitle

\section{Introduction}

The evolution of collisionless plasmas in the presence of electromagnetic
fields is well described by the Vlasov-Maxwell set of equations. The
Vlasov equation describes the evolution of the probability distribution
functions (PDFs) for one or more species, while electromagnetic fields
evolve according to Maxwell's equations. The field-PDF description
is tightly coupled. Maxwell's equations (or a subset thereof) are
driven by moments of the PDF such as charge density and/or current
density. The PDF, on the other hand, follows a hyperbolic equation
in phase space, whose characteristics are self-consistently determined
by the fields. 

The numerical solution of such strongly coupled systems has proved
challenging. Here, we concern ourselves with particle-in-cell (PIC)
techniques \citep{birdsall-langdon,hockneyeastwood,grigoryev2002numerical},
which integrate the Vlasov-Maxwell equations in time by combining
a mesh (to represent fields) with particles (to follow characteristics
of the PDF in phase space). The PIC approach has been very successful,
enabling many first-principle kinetic calculations of plasma dynamics
since the early years of computer simulations (e.g. \citep{dawson1960plasma}).
Key aspects of the method involve the definition of interpolation
operations between mesh and particle quantities, and the coupled temporal
integration of field and particle equations.

The most common PIC implementation employs an explicit time integration
of the field and particle equations \citep{birdsall-langdon} (e.g.,
the full-\textit{f} explicit momentum-conserving method). Explicit
approaches provide a straightforward recipe for temporal integration.
However, standard explicit PIC approaches suffer from temporal numerical
stability constraints (the well-known Courant-Friedrichs-Lewy or CFL
time step limit). Furthermore, some explicit PIC implementations also
feature spatial stability constraints (e.g., the momentum-conserving
explicit Vlasov-Poisson PIC formulation, which requires the Debye
length to be resolved in order to avoid so-called finite-grid instabilities)
\citep{birdsall-langdon}. As a result of these stability constraints,
explicit PIC becomes very demanding computationally when applied to
multidimensional configurations in the presence of general electromagnetic
fields\citep{bowers-pop-08-vpic}.

Implicit methods, however, can free the PIC approach from numerical
stability constraints, and thus have the potential of much improved
algorithmic efficiency. This realization drove the exploration of
implicit PIC starting in the 1980s \citep{mason-jcp-81-im_pic,denavit-jcp-81-im_pic,brackbill-forslund,celest1d,friedman-cppcf-81-di_pic,cohen-jcp-82-ipic,langdon-jcp-83-di_pic,barnes-jcp-83-di_pic,brackbill-mts-85,langdon1985multiple,cohen1986multiple,hewett-jcp-87-di_pic,friedman1990second,kamimura1992implicit,celest3d}.
These studies explored the viability of an implicit PIC formulation
and its accuracy properties, and resulted in important developments
such as the implicit-moment method \citep{mason-jcp-81-im_pic,denavit-jcp-81-im_pic,brackbill-forslund,brackbill-mts-85,celest1d,celest3d}
and the direct-implicit method \citep{friedman-cppcf-81-di_pic,langdon-jcp-83-di_pic,langdon1985multiple,hewett-jcp-87-di_pic,kamimura1992implicit}.
However, limitations of the solver technology at the time forced early
implicit PIC practitioners to rely on approximations such as linearization
and lagging, which did not respect the strong field-particle coupling.
These numerical approximations produced energy conservation errors
that could result in significant plasma self-heating or self-cooling
\citep{cohen-jcp-89-ipic_perf}.

In this study, we explore a fully implicit PIC solver, based on Newton-Krylov
methods \citep{kelley}, in which field-particle couplings are converged
to a tight nonlinear tolerance. For this proof-of-principle study,
we employ a one-dimensional (1D) electrostatic Vlasov-Ampère (VA)
model. This study builds on earlier studies for the 1D Vlasov-Poisson
(VP)~\citep{kim2005fullyimplicit} and VA~\citep{chen2010fullyimplicit}
equations. The emphasis here is on \emph{both} accuracy and efficiency.
By iterating particles and fields to a tight convergence tolerance,
the approach features superior stability and accuracy properties,
avoiding most of the accuracy pitfalls in earlier implicit PIC implementations.
In particular, we will show that the formulation is stable against
both temporal and spatial instabilities. The formulation is charge-
and energy-conserving to numerical round-off for arbitrary implicit
time steps. This is unlike the explicit {}``energy-conserving''
PIC formulation of Lewis \citep{lewis1970energy}, which conserves
energy only in the limit of arbitrarily small time steps. Automatic
exact charge conservation\citep{villasenor1992rigorous} is achieved
by ensuring that each particle moves within a given cell. Momentum
is not exactly conserved in our approach, but errors are kept small
by consistent spatial smoothing and an adaptive particle sub-stepping
orbit integrator, which is instrumental to prevent particle tunneling
(a deleterious effect for long-term accuracy). The time-centered VA
equations are orbit-averaged along the sub-stepped particle orbits
\citep{cohen-jcp-82-orbit_averaging} to enable an exact energy conservation
theorem. As a result, very large time steps, constrained only by the
collective dynamical time scale of interest, are possible for the
field solver without significant accuracy loss. 

Algorithmically, the approach features an unpreconditioned Jacobian-free
Newton-Krylov (JFNK) solver. A main algorithmic contribution of this
study is the nonlinear elimination of the new-time particle variables
(positions and velocities), which we term particle enslavement. This
results in a two-fold advantage. On one hand, the nonlinear formulation
features computer memory requirements comparable to those of a fluid
computation. On the other hand, it affords us substantial freedom
in regards to the particle orbit integrator, enabling the implementation
of an adaptive charge-conserving mover, as described earlier.

Numerical examples are presented that demonstrate the advertised properties
of the scheme. In particular, we have performed multiscale ion acoustic
wave (IAW) simulations. The numerical results show that an accurate
orbit integration is key for long-term accuracy. Furthermore, our
results demonstrate that a standard explicit VP momentum-conserving
PIC solver requires a timestep much smaller than the explicit CFL
to provide comparable accuracy to the fully implicit PIC algorithm
employing a much larger time step. Efficiency-wise, we argue that
large CPU gains are possible, especially in multiple dimensions, when
the system size is much larger than the Debye length. We demonstrate
moderate CPU gains numerically with the 1D IAW problem with our unpreconditioned
Newton-Krylov solver, underscoring the algorithmic potential of the
approach.

The rest of the paper is organized as follows. Section \ref{sec:VA-formulation-continuum}
describes the Vlasov-Ampère model in the continuum and its application
to 1D electrostatic plasma simulation. Section \ref{sec:VA-discrete-implemetation}
describes the details of our discrete Vlasov-Ampère implementation,
with emphasis on energy and charge conservation, as well as our adaptive
particle mover. Section \ref{sec:jfnk_solver} describes the concept
of particle enslavement, and provides details of our JFNK solver implementation.
Section \ref{sec:Num-Examples} demonstrates the advertised accuracy
and efficiency properties of the scheme with standard electrostatic
problems such as Landau damping, two-stream instability, and the ion
acoustic wave. Finally, we conclude in section \ref{sec:Conclusion}.

\section{Electrostatic Vlasov-Ampère model}

\label{sec:VA-formulation-continuum}

A collisionless electrostatic plasma is described by the Vlasov-Poisson
equations:\begin{eqnarray}
\frac{\partial f_{\alpha}}{\partial t}+\mathbf{v}\cdot\nabla f_{\alpha}+\frac{q_{\alpha}}{m_{\alpha}}\mathbf{E}\cdot\nabla_{v}f_{\alpha} & = & 0,\label{eq:vlasov}\\
\nabla\cdot\mathbf{E} & = & \frac{\rho}{\epsilon_{0}},\label{eq:poisson}\\
\mathbf{E} & = & -\nabla\phi,\label{eq:E}\end{eqnarray}
where $f_{\alpha}(\mathbf{r},\mathbf{v})$ is the particle distribution
function of species $\alpha$ in phase space, $q_{\alpha}$ and $m_{\alpha}$
are the species charge and mass respectively, $\phi$ and $\mathbf{E}$
are the self-consistent electric potential and field respectively,
$\rho(\mathbf{r})=\sum_{\alpha}q_{\alpha}\int d\mathbf{v}f_{\alpha}(\mathbf{r},\mathbf{v})$
is the charge density, and $\epsilon_{0}$ is the vacuum permittivity.
This system of equations is commonly used for modeling the behavior
of an electrostatic plasma in one or more dimensions~\citep{birdsall-langdon}.
An alternate formulation can be written using Ampère's law, which
is derived in the electrostatic limit as follows. We start with the
charge continuity equation:\begin{equation}
\frac{\partial\rho}{\partial t}+\nabla\cdot\mathbf{j}=0,\label{eq:continuity}\end{equation}
where $\mathbf{j}=\sum_{\alpha}q_{\alpha}\int\mathbf{v}d\mathbf{v}f_{\alpha}$
is the plasma current density. We substitute Eq.(\ref{eq:poisson})
in Eq.(\ref{eq:continuity}) to have\begin{equation}
\nabla\cdot\left(\epsilon_{0}\frac{\partial\mathbf{E}}{\partial t}+\mathbf{j}\right)=0.\label{eq:divAmpere}\end{equation}
In 1D, one finds the integral of Eq.(\ref{eq:divAmpere}) exactly
as \begin{equation}
\epsilon_{0}\frac{\partial E}{\partial t}+j=C(t)\label{eq:Amperevec}\end{equation}
where $C(t)$ is an arbitrary function of time. By integrating Eq.(\ref{eq:Amperevec})
over the periodic domain, one finds $C(t)=\left\langle j\right\rangle $,
where $\left\langle j\right\rangle =\int jdx/\int dx$ is the spatial
average of the current density, and hence Eq.(\ref{eq:Amperevec})
becomes: \begin{equation}
\epsilon_{0}\frac{\partial E}{\partial t}+j=\left\langle j\right\rangle .\label{eq:Ampere1D}\end{equation}
The right hand side of Eq.(\ref{eq:Ampere1D}) is a solvability condition
for the 1D electrostatic Ampère equation. It is also required to preserve
Galilean invariance of the system. 

In the continuum, the VP and VA formulations are equivalent. In the
discrete, however, they have different properties. Generally, VP is
momentum and charge conserving, while VA can be energy and charge
conserving, as we will show. We will also show that the discrete VA
and VP can be equivalent under special conditions {[}see Sec.(\ref{sub:equivalent-VA&VP}){]}.
In what follows, we consider the VA formulation, as it generally has
better stability properties~\citep{cohen-jcp-82-orbit_averaging}.

The VA model can be extended to multi-dimensions by replacing $j-\left\langle j\right\rangle $
with the longitudinal piece of the current, $\mathbf{j}+\nabla\times(\nabla^{-2}(\nabla\times\mathbf{j})$,
where $\nabla^{-2}$ is the inverse Laplace operator. The latter expression
ensures that the electric field obtained from Ampère's law is conservative
{[}i.e., that Eq.(\ref{eq:E}) is enforced{]}. Alternatively, one
can derive an evolution equation for the electric field (or, equivalently,
the electrostatic potential $\phi$) by using the charge continuity
equation Eq.(\ref{eq:continuity}), Gauss' law Eq.(\ref{eq:poisson}),
and the electrostatic approximation Eq.(\ref{eq:E}) to find: \[
\epsilon_{0}\frac{\partial\nabla^{2}\phi}{\partial t}-\nabla\cdot\mathbf{j}=0,\]
with $\mathbf{E}$ found from Eq.(\ref{eq:E}). This equation also
requires inverting a Laplace operator to find $\phi$.

\section{Implicit particle-based discretization of the VA model}

\label{sec:VA-discrete-implemetation}

We begin to develop a suitable implicit discretization of the VA equations
by considering a time-centered {[}Crank-Nicolson (CN){]} formulation,
written as \begin{eqnarray}
\epsilon_{0}\frac{E_{i}^{n+1}-E_{i}^{n}}{\Delta t}+j_{i}^{n+1/2} & = & \left\langle j\right\rangle ^{n+1/2},\label{eq:discete-Ampere}\\
\frac{x_{p}^{n+1}-x_{p}^{n}}{\Delta t} & = & v_{p}^{n+1/2},\label{eq:discrete-xp}\\
\frac{v_{p}^{n+1}-v_{p}^{n}}{\Delta t} & = & \frac{q_{p}}{m_{p}}E(x_{p}^{n+1/2}),\label{eq:discrete-vp}\end{eqnarray}
where subscripts $i$ and $p$ denote grid points and particles respectively,
superscript $n$ denotes time step, $j_{i}^{n+1/2}=\sum_{p}q_{p}S(x_{i}-x_{p}^{n+1/2})v_{p}^{n+1/2}/\Delta x$
is the time-centered plasma current density, $E(x_{p}^{n+1/2})=\sum_{i}S(x_{i}-x_{p}^{n+1/2})(E_{i}^{n+1}+E_{i}^{n})/2$
is the time-centered electric field, and the half time step particle
position and velocity are defined by $x_{p}^{n+1/2}=(x_{p}^{n}+x_{p}^{n+1})/2$
and $v_{p}^{n+1/2}=(v_{p}^{n}+v_{p}^{n+1})/2$, respectively. The
CN method is implicit (a particle's future position and velocity depend
on the future acceleration), symplectic, second order accurate in
time, and unconditionally stable. Most importantly, CN is non-dissipative,
which enables an exact energy conservation theorem. Next, we proceed
to show that the time centered VA equations (\ref{eq:discete-Ampere})-(\ref{eq:discrete-vp})
admit a discrete energy conservation theorem.

\subsection{Exact energy conservation theorem}

Multiplying Eq.(\ref{eq:discrete-vp}) by $v_{p}^{n+1/2}$ and summing
over all particles, we find that\begin{eqnarray*}
\sum_{p}\frac{1}{2}m_{p}\left(v_{p}^{n+1}+v_{p}^{n}\right)\left(v_{p}^{n+1}-v_{p}^{n}\right) & = & \sum_{p}v_{p}^{n+1/2}q_{p}E(x_{p}^{n+1/2})\Delta t\\
\textsf{\small introducing def. of}\: E(x_{p}^{n+1/2})\rightarrow & = & \sum_{p}\Delta tv_{p}^{n+1/2}q_{p}\sum_{i}\frac{E_{i}^{n+1}+E_{i}^{n}}{2}S(x_{i}-x_{p}^{n+1/2})\\
\textsf{\small commuting sums}\rightarrow & = & \sum_{i}\Delta tj_{i}^{n+1/2}\frac{E_{i}^{n+1}+E_{i}^{n}}{2}\Delta x\\
\textsf{\small using Ampère's law}\rightarrow & = & \sum_{i}\left[\Delta t\left\langle j\right\rangle ^{n+1/2}-\epsilon_{0}\left(E_{i}^{n+1}-E_{i}^{n}\right)\right]\frac{E_{i}^{n+1}+E_{i}^{n}}{2}\Delta x\\
\sum_{i}E_{i}^{n(+1)}=0\rightarrow & = & -\sum_{i}\frac{\epsilon_{0}}{2}\left[\left(E_{i}^{n+1}\right)^{2}-\left(E_{i}^{n}\right)^{2}\right]\Delta x\end{eqnarray*}
where the last step is justified because $E$ is a gradient of $\phi$
and we are considering a periodic domain. We therefore conclude that
\begin{equation}
\left[\sum_{p}\frac{1}{2}m_{p}\left(v_{p}\right)^{2}+\sum_{i}\frac{\Delta x}{2}\epsilon_{0}\left(E_{i}\right)^{2}\right]^{n+1}=\left[\sum_{p}\frac{1}{2}m_{p}\left(v_{p}\right)^{2}+\sum_{i}\frac{\Delta x}{2}\epsilon_{0}\left(E_{i}\right)^{2}\right]^{n},\label{eq:en-conservation-cn}\end{equation}
i.e., the electrostatic system (\ref{eq:discete-Ampere})-(\ref{eq:discrete-vp})
features an exact energy conservation theorem, whereby the total energy
{[}defined as the sum of kinetic energy ($\sum_{p}\frac{1}{2}mv_{p}^{2}$)
and the electric field energy ($\sum_{i}\frac{\Delta x}{2}\epsilon_{0}E_{i}^{2}$){]}
is exactly conserved from time step $n$ to time step $n+1$. We emphasize
that this result is enabled by a VA formulation that uses (a) a CN
time discretization and (b) identical current assignment and force
interpolation shape functions. It is important to note that exact
conservation of energy requires the implicit field and particle equations
to be updated in a nonlinearly consistent manner every time step.
This is distinctly different from previous implicit schemes~\citep{langdon1985implicit,cohen1986multiple,drouin2010particle},
which do not feature nonlinear consistency. It is also different from
the {}``energy-conserving'' scheme developed by Lewis~\citep{lewis1970energy},
which is explicit and does not conserves energy exactly with finite
$\Delta t$. 

The benefit of the existence of an energy conservation theorem is
that finite-grid instabilities, commonly seen in explicit PIC simulations
when $\Delta x>\lambda_{D}$, are eliminated. This property has the
potential of significant computational savings of implicit PIC vs.
explicit PIC, as many fewer grid points (and consequently fewer particles)
will be required for the former in regimes where $k\lambda_{D}\ll1$.
Additionally, time steps much larger than $\Delta t_{CFL}\sim\omega_{pe}^{-1}$
are possible without numerical instabilities or numerical heating
or cooling. However, stability is not the only requirement, as accuracy
also needs to be ensured, particularly when employing large time steps.
In the next section, we discuss the issue of accuracy in our implicit
PIC implementation, and our approach to mitigate potential numerical
errors.

\subsection{Particle sub-stepping and orbit-averaging}

\label{sub:pcle-substepping-orbavg}

In multiple time-scale problems, $\omega\ll\omega_{pe}$, and a large
time step is desired. However, as pointed out by Langdon~\citep{langdon-jcp-79-pic_ts,langdon1985implicit},
large inaccuracies in particle orbits may result when a large time
step is employed, such that a particle moves a distance much greater
than a cell size. The accuracy impact of orbit errors can be clearly
seen by computing the dielectric constant of a homogeneous plasma~\citep{langdon-jcp-79-pic_ts}.
For instance, in the intermediate frequency regime $\omega_{pi}\ll\omega\ll\omega_{pe}$,
the electron response in the dielectric constant is expected from
theory to be\[
\epsilon_{\mathrm{theory}}\cong1+(k\lambda_{D})^{-2}+(\mathrm{ion\, response}).\]
 With a CN discretization of particle orbits, however, the numerical
response when $kv_{th}\Delta t\gg1$ is \[
\epsilon_{\mathrm{CN}}\cong1+\frac{1}{4}(\omega_{pe}\Delta t)^{2}+(\mathrm{ion\, response}),\]
which is much larger than $\epsilon_{\mathrm{theory}}$ in the limit
considered. Thus, accuracy requires that \begin{equation}
\Delta t<\frac{\Delta x}{v_{th}}.\label{eq:dt-accuracylimit}\end{equation}
This is an undesirable constraint, especially for multiscale problems
for which the time-scale of the slow dynamics is much larger than
that determined by Eq.(\ref{eq:dt-accuracylimit}). 

In this study, we address the issue of orbit accuracy by considering
particle sub-stepping~\citep{cohen-jcp-82-orbit_averaging}, where
different time steps for orbit and field integration are employed:
small steps ($\Delta\tau$ such that $k_{max}v_{th}\Delta\tau<1$
with $k_{max}=\pi/\Delta x$) are used to ensure the accuracy of the
particle orbits and the collective response; meanwhile, a larger time
step ($\Delta t\geq\Delta\tau$) is used in the field solver to follow
the low-frequency dynamics. This approach allows slow particles to
be integrated quickly and fast particles to be integrated accurately.
We will discuss the numerical implementation of particle sub-stepping
in the overall algorithm in Sec.(\ref{sec:jfnk_solver}).

In principle, particle sub-stepping breaks the energy conservation
theorem. However, energy conservation can be recovered by considering
an appropriate orbit-averaged plasma current density to advance $E$
in Eq.(\ref{eq:discete-Ampere}). The appropriate time-centered, orbit
averaged current density is defined as\begin{equation}
\overline{j}_{i}^{n+1/2}=\frac{1}{\Delta x\Delta t}\sum_{\nu=1}^{N_{\nu}}\sum_{p}q_{p}S(x_{i}-x_{p}^{\nu+1/2})v_{p}^{\nu+1/2}\Delta\tau^{\nu},\label{eq:javerage}\end{equation}
where $\nu$ is the a sub-step index, $N_{\nu}$ is the number of
sub-steps, and $\Delta t=\sum_{\nu=1}^{N_{\nu}}\Delta\tau^{\nu}$.
Each sub-step is a CN move with time step $\Delta\tau^{\nu}$, which
can change from one sub-step to another:\begin{eqnarray}
\frac{x_{p}^{\nu+1}-x_{p}^{\nu}}{\Delta\tau^{\nu}} & = & v_{p}^{\nu+1/2},\label{eq:xp-nu}\\
\frac{v_{p}^{\nu+1}-v_{p}^{\nu}}{\Delta\tau^{\nu}} & = & \frac{q_{p}}{m_{p}}E(x_{p}^{\nu+1/2}),\label{eq:vp-nu}\end{eqnarray}
with $E(x_{p}^{\nu+1/2})=\sum_{i}S(x_{i}-x_{p}^{\nu+1/2})(E_{i}^{n+1}+E_{i}^{n})/2$.
Note that the $E$-field used to move the particles is averaged over
the macro-steps ($n$ and $n+1$), which is consistent with the assumption
that the electric field evolves in a slower time scale. Using $\overline{j}_{i}^{n+1/2}$
in Eq.(\ref{eq:discete-Ampere}), the energy conservation theorem
is recovered as follows:

\begin{eqnarray}
\sum_{p}\frac{1}{2}m_{p}\left[(v_{p}^{n+1})^{2}-(v_{p}^{n})^{2}\right] & = & \sum_{p}\sum_{\nu}\frac{1}{2}m_{p}\left[(v_{p}^{\nu+1})^{2}-(v_{p}^{\nu})^{2}\right]\nonumber \\
 & = & \sum_{p}\sum_{\nu}\Delta\tau^{\nu}v_{p}^{\nu+1/2}q_{p}E(x_{p}^{\nu+1/2})\nonumber \\
 & = & \sum_{p}\sum_{\nu}\Delta\tau^{\nu}v_{p}^{\nu+1/2}q_{p}\sum_{i}\frac{E_{i}^{n+1}+E_{i}^{n}}{2}S(x_{i}-x_{p}^{\nu+1/2})\nonumber \\
 & = & \sum_{i}\Delta t\overline{j}_{i}^{n+1/2}\frac{E_{i}^{n+1}+E_{i}^{n}}{2}\Delta x\nonumber \\
 & = & \sum_{i}\frac{\Delta x}{2}\epsilon_{0}\left[(E_{i}^{n+1})^{2}-(E_{i}^{n})^{2}\right].\label{eq:orbitavg-energyconservation}\end{eqnarray}

\subsection{Charge conservation}

\label{sub:charge-conservation}

Standard methods of current and charge assignment in VA do not satisfy
the continuity equation (\ref{eq:continuity})~\citep{birdsall-langdon}.
Accordingly, charge is not conserved locally, and Gauss's law is violated.
Methods for automatic, exact charge conservation for PIC simulations
have been developed \citep{Buneman1969fast,morse1971numerical,villasenor1992rigorous,esirkepov2001exact}.
In general, the charge-conserving schemes are valid only within a
given cell, and break when a particle crosses a cell boundary. The
problem is usually resolved~\citep{villasenor1992rigorous} by splitting
current contributions between adjacent cells, as follows:\[
j_{p}=j_{p1}+j_{p2},\]
 where $j_{p}=q_{p}\Delta x_{p}/\Delta t$, $j_{p1}=q_{p}\Delta x_{p1}/\Delta t$,
$j_{p2}=q_{p}\Delta x_{p2}/\Delta t$, $\Delta x_{p1}$ and $\Delta x_{p2}$
are segments of the particle displacement within a cell, and $\Delta x_{p1}+\Delta x_{p2}=\Delta x_{p}$.
By doing so, charge is strictly conserved. However, in our context,
this approach would break energy conservation, because it requires
the time-centered, orbit-averaged current density shown in the preceding
section. Here, we pursue an alternative way of assigning charge-conserving
currents, which is to actually make the particle stop at each cell
boundary. It can be readily shown (see Appendix~\ref{sec:charge-conserv-1d})
that this prescription leads to exact charge conservation for \emph{each}
particle sub-step. The property also holds in an orbit-averaged sense.
This can be readily seen by taking the orbit-average {[}$\frac{1}{\Delta t}\sum_{\nu=1}^{N_{\nu}}\Delta\tau^{\nu}${]}
of the continuity equation for each sub-step ($\nu\rightarrow\nu+1$):\begin{equation}
\frac{\rho_{i+1/2}^{\nu+1}-\rho_{i+1/2}^{\nu}}{\Delta\tau^{\nu}}+\frac{j_{i+1}^{\nu+1/2}-j_{i}^{\nu+1/2}}{\Delta x}=0,\label{eq:continuity-nu}\end{equation}
and using Eq.(\ref{eq:javerage}) and that $(\rho_{i+1/2}^{n+1}-\rho_{i+1/2}^{n})=\sum_{\nu}(\rho_{i+1/2}^{\nu+1}-\rho_{i+1/2}^{\nu})$,
to find:\begin{equation}
\frac{\rho_{i+1/2}^{n+1}-\rho_{i+1/2}^{n}}{\Delta t}+\frac{\bar{j}_{i+1}^{n+1/2}-\bar{j}_{i}^{n+1/2}}{\Delta x}=0.\label{eq:charg-conserv-orbavg}\end{equation}
This is the continuity equation in terms of the orbit-averaged plasma
current. It follows that, once appropriate particle sub-stepping and
orbit-averaging schemes are used, our VA implementation simultaneously
conserves energy and charge exactly. Unlike the exact energy conservation,
charge conservation is ensured regardless of nonlinear convergence
tolerance. This is so because the property is built into the particle
mover.

\subsection{Particle tunneling and adaptive orbit integration }

\label{sub:tunnelling-substepping}

\begin{figure}[t]
\begin{centering}
\includegraphics{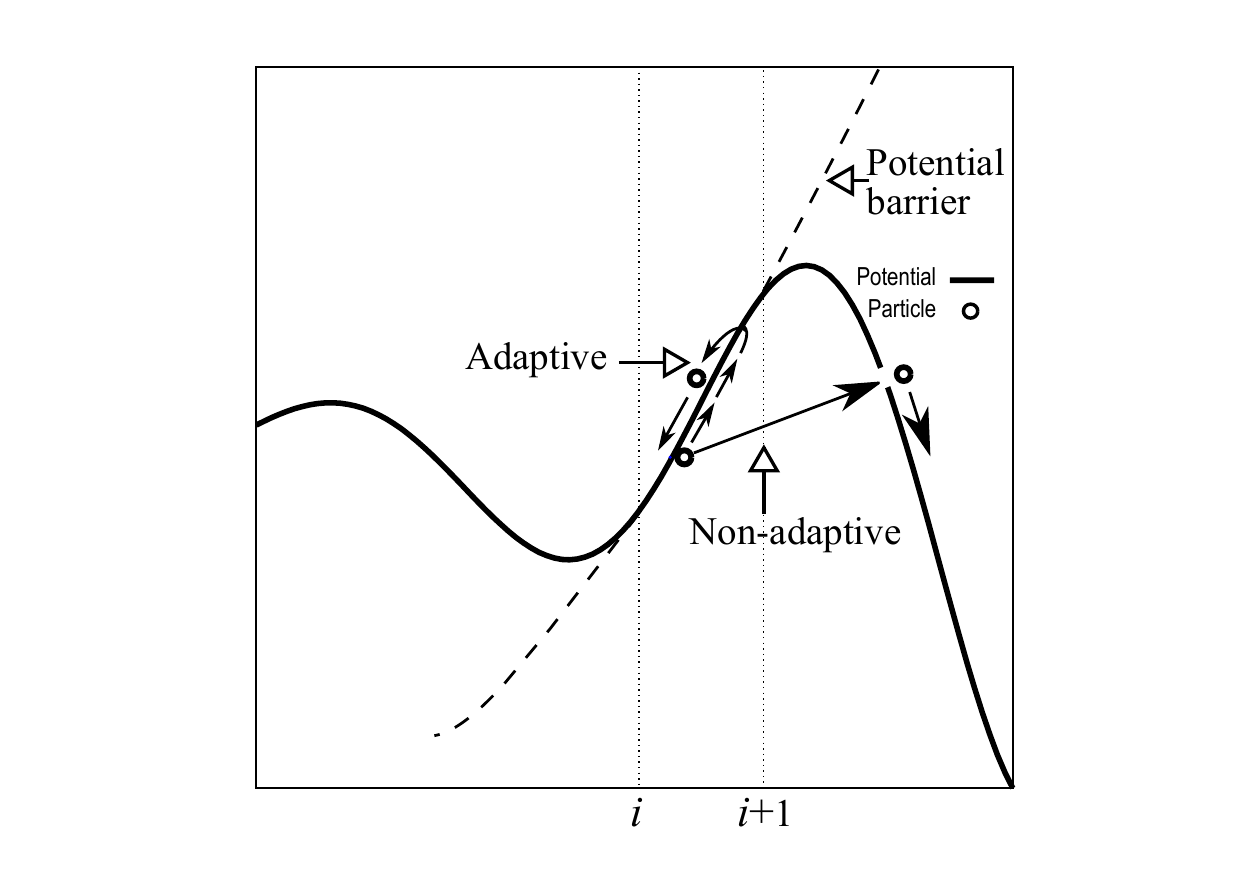}
\par\end{centering}

\caption{\label{Flo:tunneling} Illustration of particle tunneling through
a potential barrier. With charge-conserving, adaptive stepping, a
trapped particle is not allowed to tunnel through the potential barrier. }

\end{figure}

The energy- and change-conserving particle mover does not enforce
exact momentum conservation. We have found that momentum conservation
errors are exacerbated by particle tunneling. This occurs when a particle
tunnels through a potential energy barrier, which is statistically
unavoidable with a mover employing a fixed time step (see Fig.\ref{Flo:tunneling}).
The orbit error caused by particle tunneling affects momentum conservation,
because the particle ends up traveling in the wrong direction. While
only a few particles experience tunneling, we have found its accuracy
impact to be deleterious {[}This will be demonstrated numerically
in Sec.(\ref{sec:Num-Examples}){]}.

Particle tunneling is avoided by our charge-conserving particle mover,
as particles are forced to stop at cell boundaries. To further improve
the accuracy of our orbit integrator, we have designed an adaptive
orbit-integration algorithm. We wish to control the sub-time step
$\Delta\tau$ of a particle such that the local truncation error of
Eqs.(\ref{eq:xp-nu}) and (\ref{eq:vp-nu}) is below a specified tolerance.
This leads to the condition: \begin{equation}
\left\Vert le(\Delta\tau)\right\Vert _{2}<\varepsilon_{a}+\varepsilon_{r}\left\Vert r^{0}(\Delta\tau)\right\Vert _{2},\label{eq:localerror}\end{equation}
where $\left\Vert \cdot\right\Vert _{2}$ denotes the $L_{2}$-norm
of enclosed vector, $le(\Delta\tau)=\frac{(\Delta\tau)^{2}}{2}\{a_{p}^{\nu},\left(\frac{\partial a_{p}}{\partial x}v_{p}\right)^{\nu}\}$
is a measure of the local truncation error of the discrete orbit equations
(see Appendix \ref{sec:le-estimate}), $\varepsilon_{a}$ and $\varepsilon_{r}$
are absolute and relative tolerances, respectively, and $r^{0}(\Delta\tau)\equiv\{v^{0},a^{0}\}\Delta\tau$
is the initial residual. The key to the effectiveness of the approach
is in the estimation of $\partial a_{p}/\partial x$ in $le(\Delta\tau)$.
Rather than computing a local estimate based on the current particle
position, we compute a cell-based estimate $\partial a_{p}/\partial x\approx\frac{q}{m}(E_{i+1}-E_{i})/\Delta x$
(where we have assumed that the particle is located between grid points
$i$ and $i+1$ at time level $\nu$), and analytically continuate
this estimate beyond the cell boundaries. This provides a numerical
potential barrier for trapped particles against tunneling, as illustrated
in Fig.\ref{Flo:tunneling}. With this estimate, Eq.(\ref{eq:localerror})
results in a quadratic equation for an upper limit for $\Delta\tau$.
As before, particles stop at cell boundaries to enforce charge conservation.%

\subsection{Space filtering}

\label{sub:binomial}

In the conventional momentum-conserving scheme, finite-grid instabilities
result in large energy conservation errors~\citep{birdsall-langdon}.
In energy-conserving schemes, aliasing errors result in the loss of
momentum conservation. The aliasing errors arise as the wavelengths
of the particle moments, which are sampled in continuous space, can
be much shorter than those of moments defined on the mesh. Because
of the finite number of grid points, mode perturbations with $k\Delta x>\pi$
contribute to the modes with $k\Delta x<\pi$. The particle shape
function can be viewed as a low pass filter to reduce aliasing errors:
a B-spline of order $m$ has the form of $S_{m}(k)=\left(\sin\frac{1}{2}k\Delta x/\frac{1}{2}k\Delta x\right)^{m}$,
which depresses short-wave-length signals. One can achieve further
filtering by increasing $m$, but this has disadvantages (for instance,
it results in extended stencils which complicate a parallel implementation).
Alternatively, one can apply space filtering of grid quantities, which
is recommended to further reduce noise signals with $k\Delta x\rightarrow\pi$
\citep{birdsall1980plasma}. One simple example is the $\cos^{2}\left(\frac{k\Delta x}{2}\right)$
(binomial) filter in Fourier space. In real space, it is equivalent
to replacing a grid quantity $Q_{i}$ with a binomial operator $SM$
such that \begin{equation}
SM(Q)_{i}=\frac{Q_{i-1}+2Q_{i}+Q_{i+1}}{4}.\label{eq:bi-nomial-smoothing}\end{equation}
This is the approach followed in this study. Implementation-wise,
two extra operations are introduced in the implicit algorithm. Firstly,
the binomial operator is used on the electric field on the mesh, which
is then interpolated to the particles. The corresponding particle
motion per sub-step is given by\begin{eqnarray}
\frac{x_{p}^{\nu+1}-x_{p}^{\nu}}{\Delta\tau^{\nu}} & = & v_{p}^{\nu+1/2},\label{eq:xp-nu-sm}\\
\frac{v_{p}^{\nu+1}-v_{p}^{\nu}}{\Delta\tau^{\nu}} & = & \frac{q_{p}}{m_{p}}SM(E)(x_{p}^{\nu+1/2}).\label{eq:vp-nu-sm}\end{eqnarray}
Secondly, the binomial operator is used on the orbit-averaged current
density in Ampère's equation:\begin{equation}
\epsilon_{0}\frac{E_{i}^{n+1}-E_{i}^{n}}{\Delta t}+SM(\bar{j})_{i}^{n+1/2}=\left\langle \overline{j}\right\rangle ^{n+1/2}.\label{eq:bi-amperelaw}\end{equation}
Note that the smoothing does not change the spacial average of $\overline{j}$
on the right-hand-side. It is important to note that both the energy
and charge conservation properties still hold with binomial smoothing:
one simply needs to realize that, in a periodic domain, $\sum_{i}\overline{j}_{i}SM(E)_{i}=\sum_{i}SM(\overline{j})_{i}E_{i}$,
which allows the proof in Eq.(\ref{eq:orbitavg-energyconservation})
and Eq.(\ref{eq:continuity-nu})-(\ref{eq:charg-conserv-orbavg})
to follow through.

\subsection{Equivalence of VA and VP}

\label{sub:equivalent-VA&VP}

From the discussion above, it is clear that both Ampere's equation
and the charge continuity equation are strictly satisfied. It is therefore
easy to deduce that Gauss's law, if satisfied initially, is also strictly
satisfied at all times. This follows trivially from the following:\[
\left.\begin{array}{c}
\epsilon_{0}\frac{E_{i}^{n+1}-E_{i}^{n}}{\Delta t}+\bar{j}_{i}^{n+1/2}=\left\langle \overline{j}\right\rangle ^{n+1/2}\\
\frac{\rho_{i+1/2}^{n+1}-\rho_{i+1/2}^{n}}{\Delta t}+\frac{\bar{j}_{i+1}^{n+1/2}-\bar{j}_{i}^{n+1/2}}{\Delta x}=0\end{array}\right\} \Rightarrow\left(\frac{E_{i+1}-E_{i}}{\Delta x}-\rho_{i+1/2}\right)^{n+1}=\left(\frac{E_{i+1}-E_{i}}{\Delta x}-\rho_{i+1/2}\right)^{n}.\]
The discrete equivalence of the VA system and VP system is thus evident.
Consequently, the implicit energy- and charge-conserving scheme proposed
here can also be viewed as an energy-conserving implicit VP formulation,
and therefore shares a strong connection to Lewis's explicit energy-conserving
scheme\citep{lewis1970energy}. Both methods solve the same system
of equations and feature exact charge conservation. Both schemes define
$E$ and $\rho$ half grid apart, and the interpolation schemes are
identical, namely, the shape function (B-spline) of $E$ is one order
lower than that of $\rho$. 

There are, however, important differences between the two energy-conserving
methods. While Lewis's is explicit and energy-conserving only when
$\Delta t\rightarrow0$, ours is implicit and exactly energy-conserving
for arbitrary $\Delta t$. Moreover, unlike explicit PIC schemes (including
Lewis's), in which the time step is fixed and usually the same for
all particle and field equations, our implicit approach allows different
time steps to be used for particles vs. fields. This enables complete
flexibility in the orbit integration stage, which allows momentum
error control {[}see Sec.(\ref{sub:tunnelling-substepping}){]}.

\section{Formulation of nonlinear residual: nonlinear elimination/particle
enslavement}

\label{sec:jfnk_solver}

The goal of this study is to find the nonlinear root of the system
of equations (\ref{eq:xp-nu-sm})-(\ref{eq:bi-amperelaw}). This nonlinearly
coupled set of equations can be conceptually formulated as a nonlinear
residual of the form:\begin{equation}
\mathbf{F}(\mathbf{E},\boldsymbol{\xi})=0.\label{eq:totalResidualG}\end{equation}
Here, $\mathbf{E}=\{E_{i}\}$ with $i$ representing mesh points,
and $\boldsymbol{\xi}=\{x_{p},v_{p}\}$ with $p$ representing particles.
Upon convergence, $\mathbf{E}=\{E_{i}^{n+1}\}$, and $\boldsymbol{\xi}=\{x_{p}^{n+1},v_{p}^{n+1}\}$,
i.e., the new-time field and particle quantities. In formulating this
residual, we have naively taken both particle and field variables
as dependent variables. While in principle correct, this formulation
has one fundamental limitation, namely, that the residual has inherited
the full dimensionality of the kinetic problem. For any iterative
solution method, this will result in exceedingly large storage requirements.
Given that memory is the principal bottleneck in current and future
massively parallel computers, pursuing the formulation in Eq.(\ref{eq:totalResidualG})
does not appear promising.

There is, however, an alternative, which is to nonlinearly eliminate
the kinetic component to the low-dimensional description in terms
of fields and moments. The concept of nonlinear elimination is akin
to the well-known elimination process in linear equations, and can
be explained as follows. Let us consider a nonlinear residual $\mathbf{F}(\mathbf{X}_{1},\mathbf{X}_{2})$,
with $\mathbf{X}_{1}$, $\mathbf{X}_{2}$ two sets of dependent variables.
Let us also divide the global residual into two sets of equations,
$\mathbf{F}_{1}(\mathbf{X}_{1},\mathbf{X}_{2})=0$ and $\mathbf{F}_{2}(\mathbf{X}_{1},\mathbf{X}_{2})=0$,
and assume that $\mathbf{F}_{2}(\mathbf{X}_{1},\mathbf{X}_{2})$ can
be straightforwardly written in an explicit form as $\mathbf{X}_{2}=\mathbf{f}_{2}(\mathbf{X}_{1})$.
It follows that a new residual can be formulated as:\begin{equation}
\mathbf{F}_{1}(\mathbf{X}_{1},\mathbf{f}_{2}(\mathbf{X}_{1}))=\mathbf{G}(\mathbf{X}_{1})=0.\label{eq:Gtilde}\end{equation}
By construction, the new residual $\mathbf{G}$ has lower dimensionality
than the original one $\mathbf{F}$ (and thus requires less storage
for the Krylov and Newton iterations), but has the \emph{same} nonlinear
solution. One successful example of the use of nonlinear elimination
to significantly reduce the dimensionality of the nonlinear residual
in a nontrivial application (the solution of the Fokker-Planck transport
equation) is reported in Ref. \citep{chacon-jcp-00-fp2}.

In our context, the key point is that the particle equations of motion
(\ref{eq:xp-nu-sm}) and (\ref{eq:vp-nu-sm}) can be expressed as
$\mathbf{X}_{2}=\mathbf{f}_{2}(\mathbf{X}_{1})$ if one considers
$\mathbf{X}_{1}=\mathbf{E}$ and $\mathbf{X}_{2}=\boldsymbol{\xi}$.
That is, for each particle and for a given $\mathbf{E}$ on the mesh,
one can explicitly solve Newton's equations for the new-time particle
position and velocity, $(x_{p}^{n+1},v_{p}^{n+1})$. This requires
a local nonlinear solve, as the particle position affects the interpolation
of the electric field. However the particle equations of motion are
not stiff (we employ substepping), and we have found that Picard's
method of successive approximations (with a nonlinear tolerance of
\textbf{$10^{-10}$}) is sufficient to solve the particle equations
of motion effectively in each sub-timestep.

As a result, the original residual Eq.(\ref{eq:totalResidualG}) can
be formally rewritten as a function of $\mathbf{E}$ only:\begin{equation}
\mathbf{G}(\mathbf{E})=0.\label{eq:G-tilde-final}\end{equation}
This is what we term \emph{particle enslavement} ( particle quantities
do not appear explicitly in the residual). It shows that \emph{the
full kinetic description can be equivalently formulated in terms of
a low-dimensional residual}, with \emph{solver} memory requirements
(e.g., nonlinear residuals, Krylov subspace vectors, etc.) comparable
to those of a moment/fluid description. (It should be noted that one
still needs to save old-time and new-time particle quantities as auxiliary
variables to compute the nonlinear residual.) The concept is not restricted
to electrostatic PIC, and can be trivially generalized to an electromagnetic
formulation with $\mathbf{X}_{1}=\{E_{i},B_{i}\}$ and $\mathbf{X}_{2}$
as above. Note that particle enslavement allows complete flexibility
in the particle integration step, $\mathbf{X}_{2}=\mathbf{f}_{2}(\mathbf{X}_{1})$.
Since this operation is segregated in the calculation of the residual
$\mathbf{G}$, it is ideally suited to exploit heterogeneous computing
architectures such as general-purpose graphics processing units (GPGPUs).
This will be explored in future work. Finally, we note that this flexibility
has enabled us to explore the advanced features of our orbit integrator,
namely, particle sub-stepping, orbit averaging, temporal adaptivity,
and charge conservation.

Finding the root of the enslaved residual Eq.(\ref{eq:G-tilde-final})
is not a simple task, because nonlinear couplings are nonlocal on
the mesh, and are very nontrivial due to the complex orbit integration/averaging
step. Jacobian-free Newton-Krylov (JFNK) methods present important
advantages for this application, which we discuss briefly below. The
motivated reader can find extensive discussions on this approach elsewhere
\citep{kelley}. Newton's method solves the nonlinear system $\mathbf{G}(\mathbf{x})=\mathbf{0}$
(with $\mathbf{x}=\mathbf{E}$) iteratively by inverting linear systems
of the form:\[
\left.\frac{\partial\mathbf{G}}{\partial\mathbf{x}}\right|^{(k)}\delta\mathbf{x}^{(k)}=-\mathbf{G}(\mathbf{x}^{(k)}),\]
with $\mathbf{x}^{(k+1)}=\mathbf{x}^{(k)}+\delta\mathbf{x}^{(k)}$,
where $(k)$ denotes the nonlinear iteration number. Nonlinear convergence
is achieved when:\begin{equation}
\left\Vert \mathbf{G}(\mathbf{x}^{(k)})\right\Vert _{2}<\epsilon_{a}+\epsilon_{r}\left\Vert \mathbf{G}(\mathbf{x}^{(0)})\right\Vert _{2}=\epsilon_{t},\label{eq-Newton-conv-tol}\end{equation}
where $\left\Vert \cdot\right\Vert _{2}$ is the $L_{2}$-norm, $\epsilon_{a}=\sqrt{N}\times10^{-15}$
(with $N$ the total number of degrees of freedom) is an absolute
tolerance to avoid converging below round-off, $\epsilon_{r}$ is
the Newton relative convergence tolerance (set to $10^{-8}$ in this
work, unless otherwise specified), and $\mathbf{G}(\mathbf{x}^{(0)})$
is the initial residual.

Such linear systems are solved iteratively with Krylov methods, which
only require matrix-vector products to proceed. Because the linear
system matrix is a Jacobian matrix, such matrix-vector products can
be implemented Jacobian-free using the Gateaux derivative:\begin{equation}
\left.\frac{\partial\mathbf{G}}{\partial\mathbf{x}}\right|^{(k)}\mathbf{v}=\lim_{\epsilon\rightarrow0}\frac{\mathbf{G}(\mathbf{x}^{(k)}+\epsilon\mathbf{v})-\mathbf{G}(\mathbf{x}^{(k)})}{\epsilon},\label{eq:gateaux}\end{equation}
where in practice a small but finite $\epsilon$ is employed (p. 79
in \citep{kelley}). Thus, the evaluation of the Jacobian-vector product
only requires the function evaluation $\mathbf{G}(\mathbf{x}^{(k)}+\epsilon\mathbf{v})$,
and there is no need to form or store the Jacobian matrix. This, in
turn, allows for a memory-efficient implementation.

An inexact Newton method \citep{inexact-newton} is used to adjust
the convergence tolerance of the Krylov method at every Newton iteration
according to the size of the current Newton residual, as follows:\begin{equation}
\left\Vert J^{(k)}\delta\mathbf{x}^{(k)}+\mathbf{G}(\mathbf{x}^{(k)})\right\Vert _{2}<\zeta^{(k)}\left\Vert \mathbf{G}(\mathbf{x}^{(k)})\right\Vert _{2}\label{eq-inexact-newton}\end{equation}
where $\zeta^{(k)}$ is the inexact Newton parameter and $J^{(k)}=\left.\frac{\partial\mathbf{G}}{\partial\mathbf{x}}\right|^{(k)}$
is the Jacobian matrix. Thus, the convergence tolerance of the Krylov
method is loose when the Newton state vector $\mathbf{x}^{(k)}$ is
far from the nonlinear solution, and tightens as $\mathbf{x}^{(k)}$
approaches the solution. Superlinear convergence rates of the inexact
Newton method are possible if the sequence of $\zeta^{(k)}$ is chosen
properly (p. 105 in \citep{kelley}). Here, we employ the prescription:\begin{eqnarray*}
\zeta^{A(k)} & = & \gamma\left(\frac{\left\Vert \mathbf{G}(\mathbf{x}^{(k)})\right\Vert _{2}}{\left\Vert \mathbf{G}(\mathbf{x}^{(k-1)})\right\Vert _{2}}\right)^{\alpha},\\
\zeta^{B(k)} & = & \min[\zeta_{max},\max(\zeta^{A(k)},\gamma\zeta^{\alpha(k-1)})],\\
\zeta^{(k)} & = & \min[\zeta_{max},\max(\zeta^{B(k)},\gamma\frac{\epsilon_{t}}{\left\Vert \mathbf{G}(\mathbf{x}^{(k)})\right\Vert _{2}})],\end{eqnarray*}
with $\alpha=1.5$ , $\gamma=0.9$, and $\zeta_{max}=0.8$. The convergence
tolerance $\epsilon_{t}$ is defined in Eq.(\ref{eq-Newton-conv-tol}).
In this prescription, the first step ensures superlinear convergence
(for $\alpha>1$), the second avoids volatile decreases in $\zeta_{k}$,
and the last avoids oversolving in the last Newton iteration. 

A further advantage of Krylov methods is that they can be preconditioned
by considering the alternate (but equivalent) systems $J^{(k)}(P^{(k)})^{-1}P^{(k)}\delta\mathbf{x}^{(k)}=-\mathbf{G}^{(k)}$
(right preconditioning) or $(P^{(k)})^{-1}J^{(k)}\delta\mathbf{x}^{(k)}=-(P^{(k)})^{-1}\mathbf{G}^{(k)}$
(left preconditioning). Such preconditioned systems can be straightforwardly
and efficiently implemented in the Krylov algorithm as two consecutive
matrix-vector products. A crucial feature of preconditioning is that,
while it can substantially improve the convergence properties of the
Krylov iteration if $(P^{(k)})^{-1}\approx(J^{(k)})^{-1}$, it does
not alter the solution of the Jacobian system upon convergence (because
the solution $\delta\mathbf{x}^{(k)}$ of the preconditioned system
is the same as that of the original system). Therefore, one can explore
suitable approximations in the preconditioner for efficiency purposes
without compromising the accuracy of the converged result.

For this study, we have employed unpreconditioned JFNK (with $P^{(k)}$
the identity operator) to demonstrate the feasibility and the accuracy
properties of our fully implicit electrostatic PIC implementation.
Despite the lack of a preconditioner, we show in the next section
that large CPU speedups over explicit approaches are still possible
in 1D, demonstrating the potential of the approach. Multidimensional
applications will require effective preconditioning, which will be
the subject of future work.

\section{Numerical examples}

\label{sec:Num-Examples}

To demonstrate the accuracy and performance of the present algorithm,
we use three standard electrostatic test cases: Langmuir wave, two-stream
instability, and ion acoustic wave. We first show that a simple CN
mover (i.e., without sub-stepping) can accurately simulate the behavior
of Langmuir waves, including cold plasma oscillations and Landau damping
in a warm plasma. We then simulate a two-stream instability, demonstrating
the effects of charge conservation and adaptive sub-stepping on momentum
conservation. In the IAW case, in which large implicit time steps
($\omega_{pe}\Delta t\gg1$) can be taken, we discuss the impact of
the particle mover on the accuracy of long-time integration, and demonstrate
that moderate CPU gains of the implicit scheme vs. the explicit scheme
in the $k\lambda_{D}\ll1$ limit are possible. 

In all the cases, we assume a homogeneous collisionless electrostatic
plasma equilibrium with some initial perturbation \begin{equation}
f_{\alpha}(x,v,t=0)=f_{\alpha0}(v)\left[1+a\cos\left(\frac{2\pi n_{h}}{L}x\right)\right]\label{eq:initf}\end{equation}
where $f_{\alpha0}$ is the initial velocity distribution of species
$\alpha$, $a$ is the perturbation level, $L$ is the domain size,
and $n_{h}$ is the mode number of the perturbation ($n_{h}=1$ by
default). The computational domain is one-dimensional, featuring a
uniform grid and periodic boundary conditions. All the temporal and
spatial quantities are normalized by the inverse plasma frequency
($1/\omega_{pe}$) and the Debye length ($\lambda_{D}$), respectively.
Other parameters are provided in specific test cases below. To verify
the simulations using the implicit algorithm, we use either analytical
solutions, e.g., linear damping or instability rates, or results from
the conventional explicit momentum-conserving VP PIC simulations.
By default, binomial spatial smoothing {[}Eq.(\ref{eq:bi-nomial-smoothing}){]}
is used; first-order B-spline is used for interpolating current density
$j$ in the implicit scheme, and second-order B-spline is used for
interpolating charge density $\rho$ in the explicit scheme.

\subsection{Langmuir wave}

\begin{figure}
\begin{centering}
\includegraphics{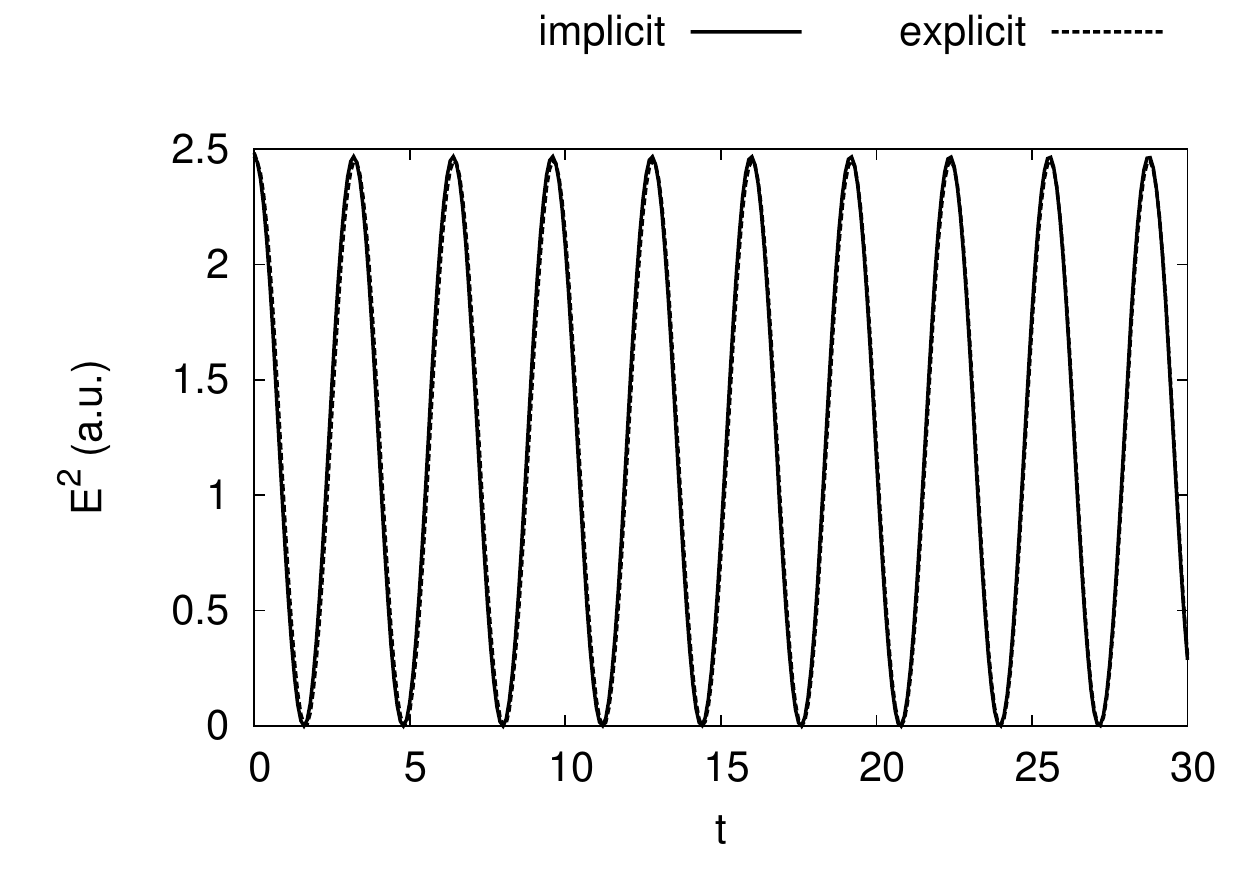}
\par\end{centering}

\caption{\label{fig:Langmuir-wave-cold}Langmuir wave in a cold plasma. Leapfrog
and simple CN are used in explicit and implicit schemes respectively.
Both schemes use a small time step ($\Delta t=0.1$). The two curves
depict the total electric field energy $E^{2}=\sum_{i}E_{i}^{2}$
and are very close to each other. The theoretical plasma frequency
$(\omega_{pe}=1)$ is accurately reproduced.}

\end{figure}

We first consider a cold plasma Langmuir wave, in which electrons
are perturbed and oscillate collectively around stationary and uniform
ions. Simulations are performed by both explicit and implicit schemes
using the same set of simulation parameters: $L=2\pi$, $N_{x}=32$,
$N_{p}=2000$, $\Delta t=0.1$, $a=0.01$, which are the domain size,
number of grid points, number of particles, the time step, and the
perturbation level {[}see Eq.(\ref{eq:initf}){]}, respectively. Figure
(\ref{fig:Langmuir-wave-cold}),shows that the leapfrog and CN schemes
agree closely with each other. In these simulations, we resolve well
the plasma period, which is the only timescale of the problem. 

\begin{figure}
\begin{centering}
\includegraphics{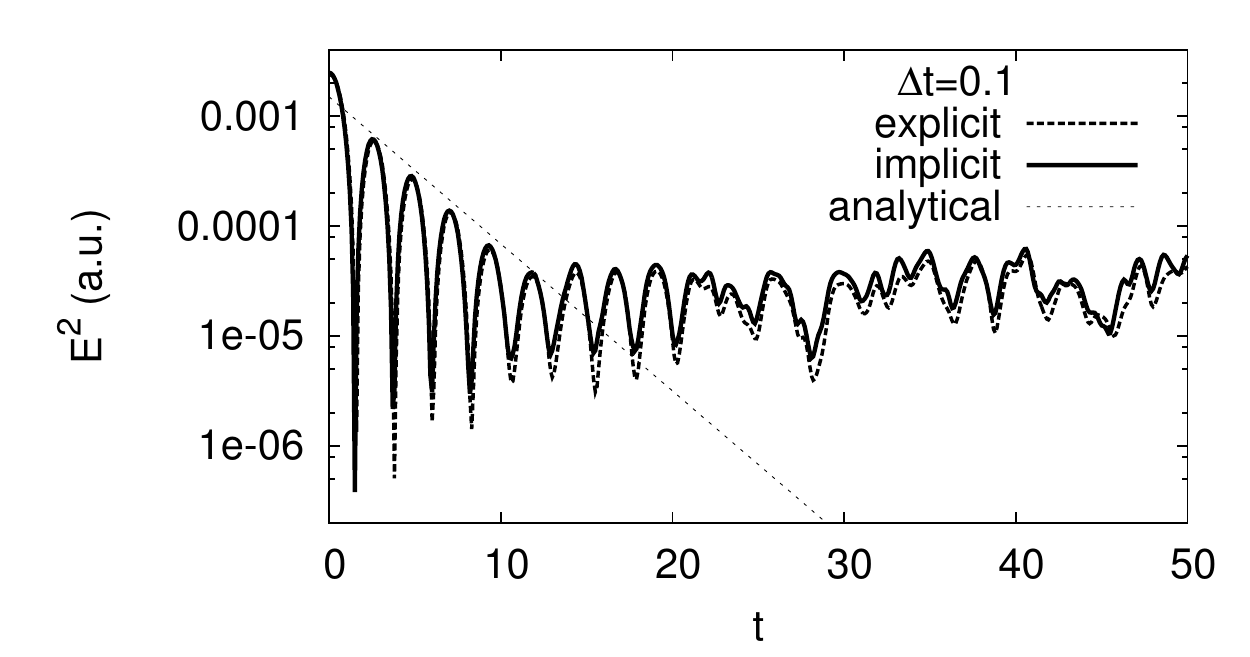}
\par\end{centering}

\begin{centering}
\includegraphics{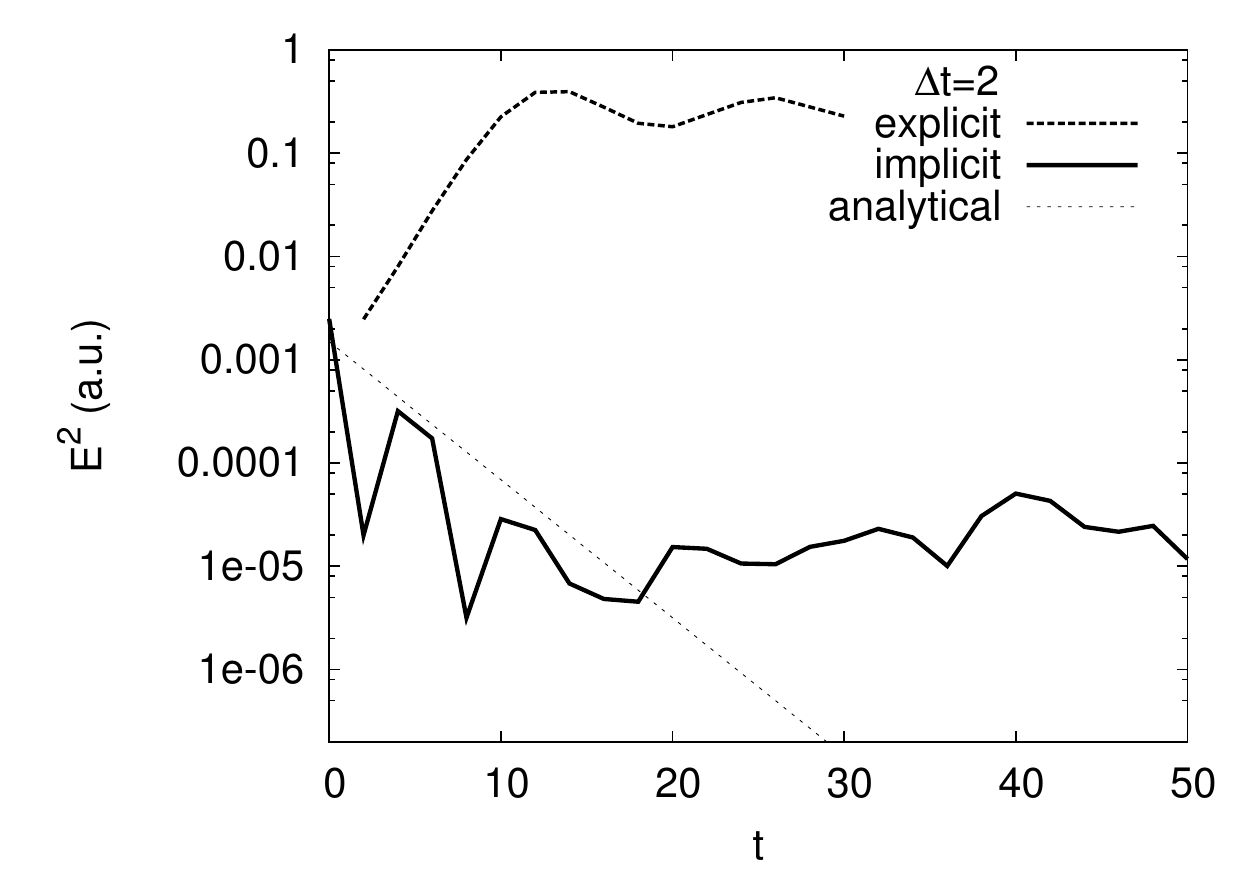}
\par\end{centering}

\caption{\label{fig:Simulated-Landau-damping}Landau damping test case. Top:
Good agreement between the explicit and the implicit schemes is seen
for $\Delta t=0.1$. Bottom: Accurate damping rate is produced by
the implicit scheme for $\Delta t=2$ (i.e., $\omega_{p}\Delta t=2$
and $\gamma\Delta t\simeq0.3$) even though the plasma frequency is
underresolved. For this time step, the explicit scheme is CFL unstable.}

\end{figure}

Next, we consider electron Landau damping of Langmuir waves in a warm
plasma, which is a pure kinetic effect. Given Maxwellian electrons
and no contribution from ions, the wave dispersion relation from linear
theory is \citep{stix1992waves}: \begin{equation}
1+\frac{1}{k^{2}}\left[1+\frac{\omega}{\sqrt{2}k}Z\left(\frac{\omega}{\sqrt{2}k}\right)\right]=0,\label{eq:plasmawaveDisper}\end{equation}
 where $Z$ is the dispersion function of Fried and Conte. Solving
Eq.(\ref{eq:plasmawaveDisper}) for a given real $k$ provides a complex
solution $(\omega+i\gamma)$, in which $\omega$ is the wave frequency
and $\gamma$ is the wave damping rate. 

We initialize the problem with a perturbation of a Maxwellian distribution
$f_{e0}$. The simulation parameters are $L=4\pi$, $N_{x}=32$, $N_{p}=40000$,
$a=0.05$, defined as before. Using $k=0.5$ in Eq.(\ref{eq:plasmawaveDisper}),
we find $\omega=1.415$ and $\gamma=0.154$. Note that Landau damping,
which mostly affects resonant particles with $v_{p}\sim\omega/k$,
occurs in a longer timescale than plasma oscillations, so that there
is a separation of timescales in this problem. We first consider a
small time step $\Delta t=0.1$, which resolves the plasma wave period.
We performed both explicit and implicit simulations under identical
conditions. Figure (\ref{fig:Simulated-Landau-damping})a compares
the results with the analytical growth rate. We see good agreement
between the two methods, and the numerical damping rate agrees well
with the theoretical one. When we increase the time step to $\Delta t=2$
{[}Fig.(\ref{fig:Simulated-Landau-damping})b{]} (such that the plasma
period is under-resolved, but the damping rate is still well resolved),
the explicit scheme is unstable due to the violation of the CFL condition.
The implicit scheme, on the other hand, is stable, and accurately
reproduces the damping rate. We note that only the simple CN mover
is employed in these implicit simulations, which enforce no local
charge conservation. The importance of local charge conservation will
be demonstrated for longer term simulations in the following sections.

\begin{figure}
\begin{centering}
\includegraphics{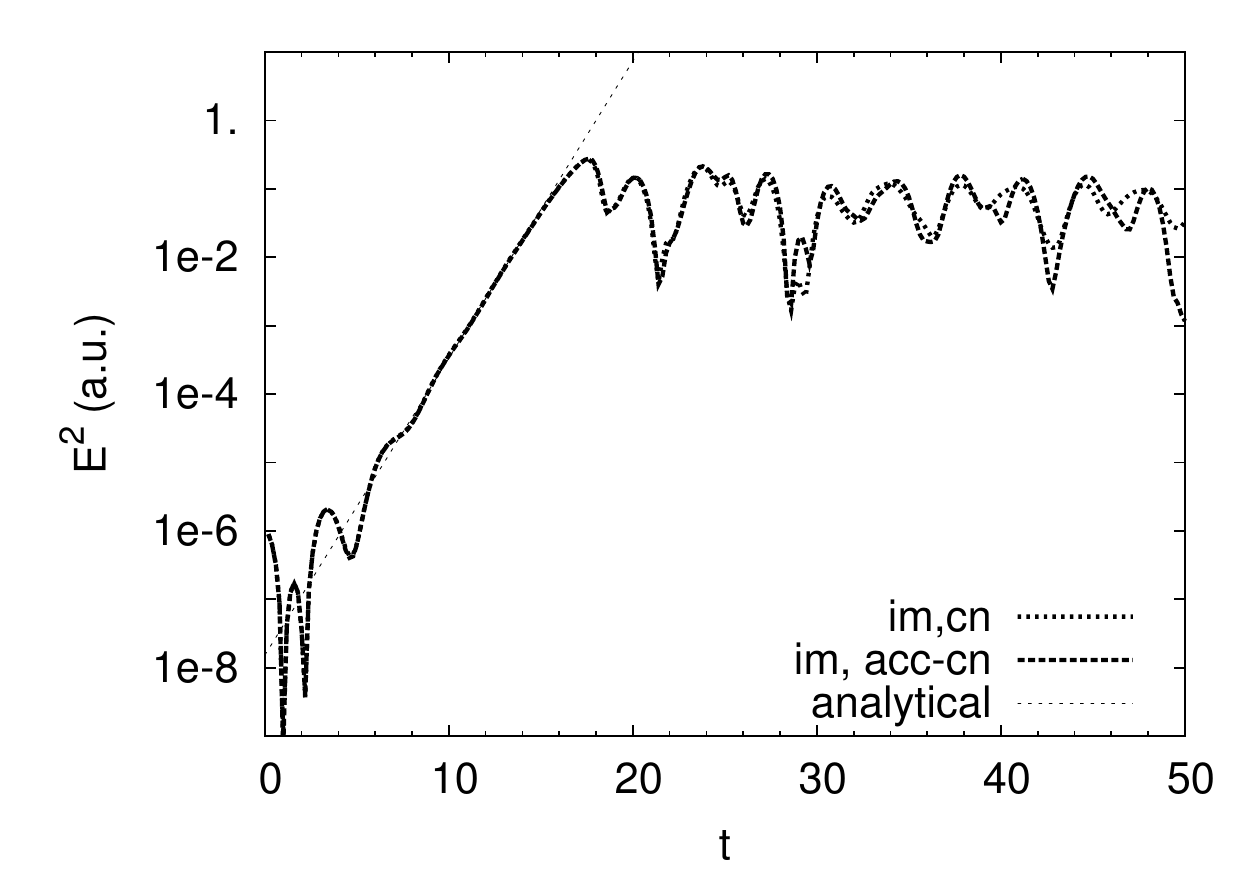}
\par\end{centering}

\caption{\label{fig:Two-stream-instability-simulated}Semi-log scale graph
of the electric field energy for the two-stream instability, obtained
with the simple CN mover and the ACC CN mover. Both simulations use
$\Delta t=0.2$. Numerical growth rates are seen to agree very well
with the analytical one ($\gamma=0.5$).}

\end{figure}

\begin{figure}
\includegraphics{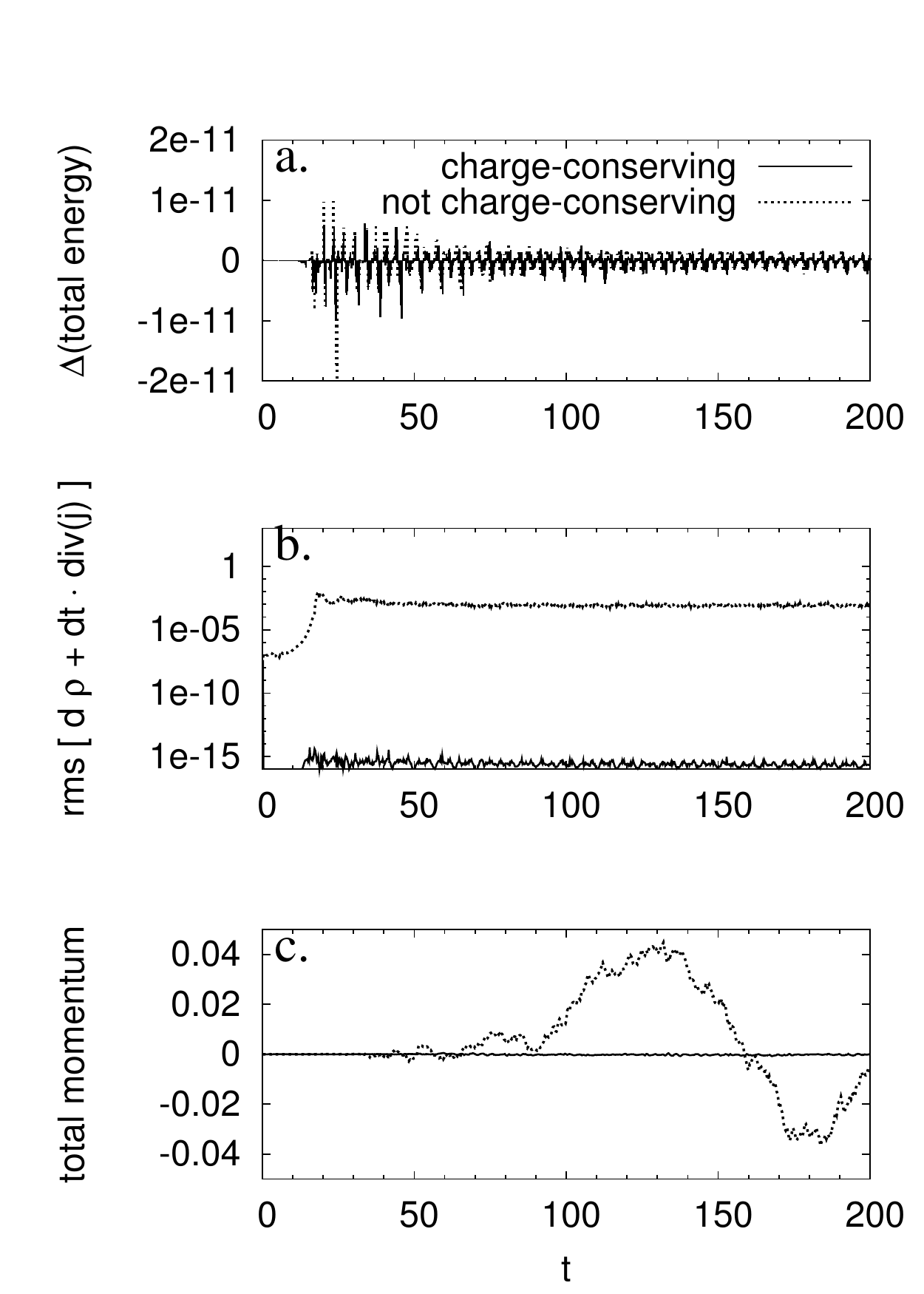}

\caption{\label{fig:compare-2stream}Comparison of the ACC CN and the simple
CN particle movers. Time histories of the change in total energy,
the root mean square of the charge continuity equation, and the total
momentum are shown in panels a,b, and c, respectively. }

\end{figure}

\subsection{Two-stream instability}

We consider two equal counter-streaming electron beams in a stationary
and uniform ion background. The analytical dispersion relation is~\citep{stix1992waves}
\begin{equation}
\frac{1}{(\omega-kv_{b})^{2}}+\frac{1}{(\omega+kv_{b})^{2}}=1,\label{eq:2streamDisper}\end{equation}
where $v_{b}$ is the beam speed. The simulation is performed for
$L=2\pi$, $v_{b}=\sqrt{3}/2$, $N_{x}=64$, $N_{p}=10000,$ $\Delta t=0.2$,
and $a=0.005$, as defined previously. Figure (\ref{fig:Two-stream-instability-simulated})
depicts the evolution of the electrostatic energy, showing that the
numerical growth rate agrees very well with the analytical one with
both the simple CN mover and the adaptive-charge-conserving (ACC)
CN mover. 

This test case is ideally suited for exploring the properties of the
ACC CN mover vs. simple CN. For this, the conserved quantities of
the system (energy, charge, and momentum) are monitored using both
CN movers. In Fig.(\ref{fig:compare-2stream})a, we see that both
CN movers conserve the total energy commensurately with the nonlinear
tolerance employed ($\epsilon_{t}=10^{-8}$), as expected. Figure
(\ref{fig:compare-2stream})b depicts the root mean square of the
discrete charge conservation equation at every cell, and confirms
that the ACC mover conserves charge exactly, while the simple CN mover
results in significant charge conservation errors. Finally, Fig. (\ref{fig:compare-2stream})c
demonstrates the substantial improvement in the conservation of total
momentum achieved by the ACC mover vs. the simple CN mover, particularly
in the nonlinear regime. 

\begin{figure}
\begin{centering}
\includegraphics[scale=0.8]{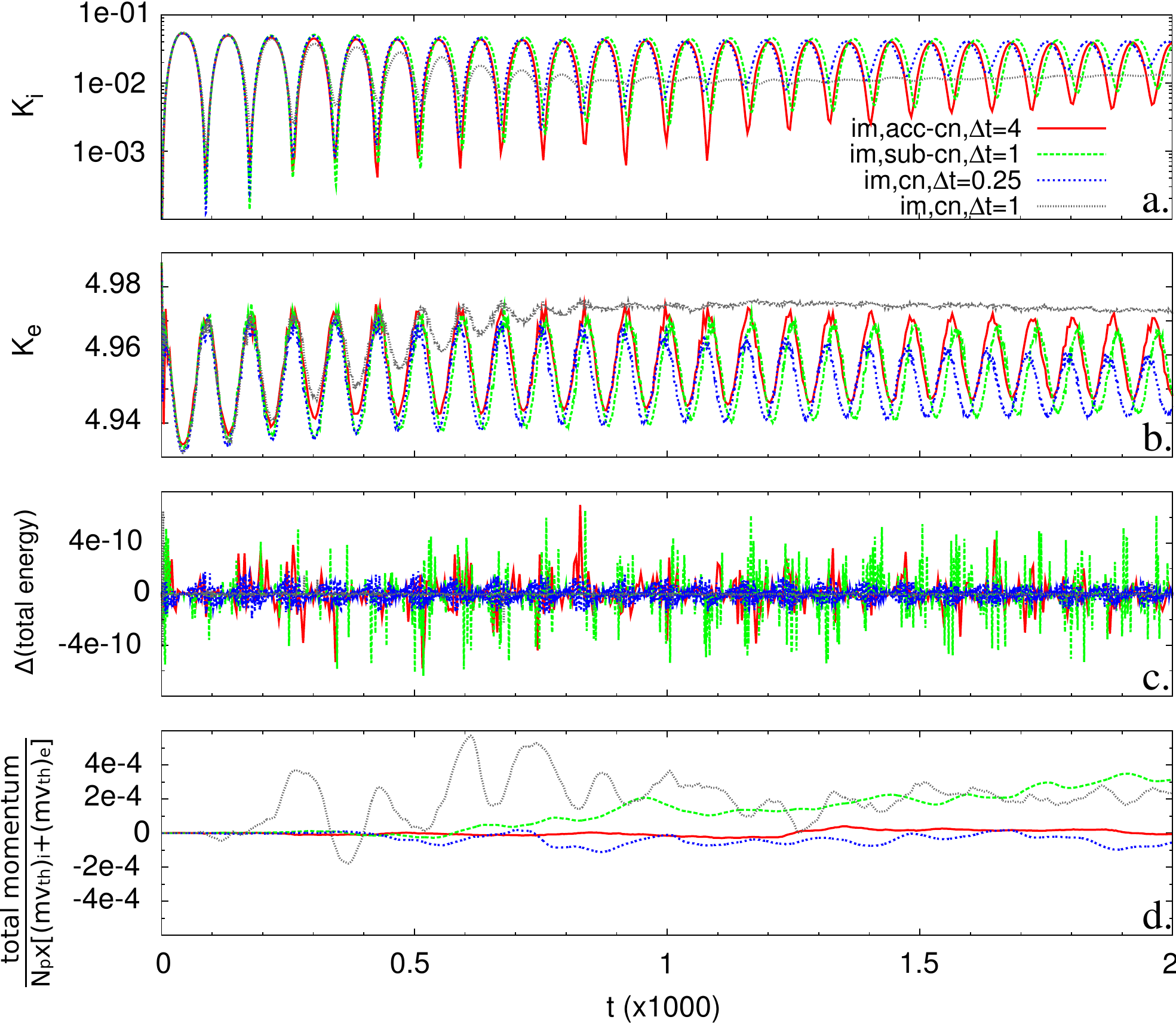}
\par\end{centering}

\caption{\label{fig:IAW-3-movers}Accuracy test of the implicit algorithm using
three movers: simple CN mover (cn), sub-stepped CN mover (sub-cn),
and adaptive-charge-conserving CN mover (acc-cn). a) Time history
of ion kinetic energy shows the impact of using fixed-time step movers
in a long-time simulation. b) Time history of electron kinetic energy
shows the importance of using small (sub-)time steps to capture the
electron dynamics, which exhibits relatively small variations. c)
Total energy is well conserved in all implicit simulations. d) Momentum
conservation is substantially improved when particle orbits are accurately
calculated using the charge-conserving self-adaptive CN mover.}

\end{figure}

\begin{figure}
\includegraphics[scale=0.8]{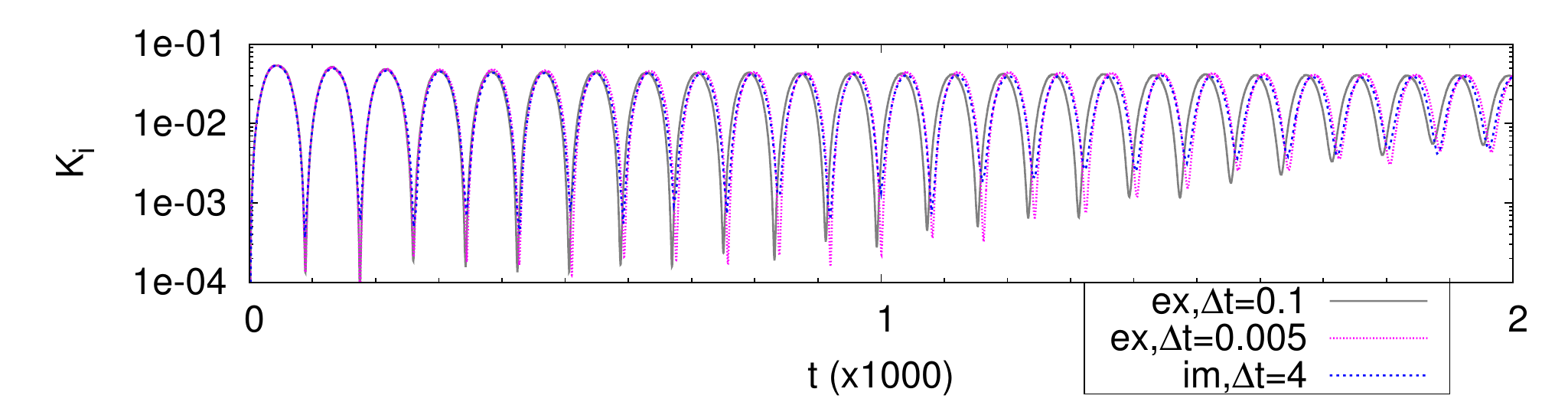}

\caption{\label{fig:IAW-ex-im}Ion kinetic energy {[}$K_{i}=\sum_{p}(\frac{1}{2}mv_{p}^{2})_{i}${]}
obtained by implicit (with ACC-CN mover) and explicit simulations. }

\end{figure}

\subsection{Ion acoustic wave}

Simulating the ion acoustic wave requires considering both electron
and ion dynamics. The wave is excited by an ion density perturbation,
and is propagated by the compression and expansion caused by the thermal
motion of ions, as well as by incomplete Debye shielding due to the
thermal motion of electrons. Adding the ion contribution to the dielectric
constant {[}cf. Eq.(\ref{eq:plasmawaveDisper}){]}, the dispersion
relation of ion acoustic wave can be written as~\citep{stix1992waves}\begin{equation}
1+\frac{1}{k^{2}}\left[1+\frac{\omega}{\sqrt{2}k}Z\left(\frac{\omega}{\sqrt{2}k}\right)\right]+\frac{\sqrt{r_{T}}}{k^{2}}\left[1+\frac{\sqrt{r_{T}r_{m}}\omega}{\sqrt{2}k}Z\left(\frac{\sqrt{r_{T}r_{m}}\omega}{\sqrt{2}k}\right)\right]=0,\label{eq:IAWDisper}\end{equation}
where $r_{T}=T_{e}/T_{i}$ is the ratio of electron and ion temperatures,
and $r_{m}=m_{i}/m_{e}$ is the ratio of ion and electron masses.
In regimes with $r_{T}\gg1$ and $r_{m}\gg1$, the IAW frequency is
much smaller than the plasma frequency ($\omega\ll\omega_{pe}$),
and the wavelength is much longer than the Debye length ($k\lambda_{D}\ll1$),
which make the IAW a truly multiscale problem. 

The main goals of this section are to examine the long-term accuracy
of the implicit algorithm employing different particle movers, and
to provide an accuracy and efficiency comparison with a standard explicit
VP algorithm with a leap-frog mover. We perform implicit and explicit
simulations under identical conditions with $r_{T}=10^{4}$, $r_{m}=200$,
$a=0.2$, $L=10$, $N_{x}=32$, $N_{p}=128000$. Note that electrons
are much faster than ions for these parameters. Therefore, both stability
and accuracy will be largely controlled by electron dynamics. 

We first consider four different implicit movers: a single-step CN
mover with $\Delta t=\Delta\tau=0.25$ and $\Delta t=\Delta\tau=1$,
a sub-stepping CN mover with $\Delta t=4\Delta\tau=1$ (i.e., 4 particle
sub-steps), and an adaptive charge-conserving CN mover. Simulation
results with these movers for IAW are depicted in Fig.(\ref{fig:IAW-3-movers}).
Figure (\ref{fig:IAW-3-movers})a depicts the evolution of the ion
kinetic energy, and motivates several important observations. Firstly,
the single-step CN with $\Delta t=1$ fails to capture the long-term
IAW behavior. This is because the average distance that an electron
travels is larger than a cell size, namely $k_{max}v_{the}\Delta t>1$.
In this regime, Debye shielding is not captured accurately, as predicted
by theory~\citep{langdon-jcp-79-pic_ts}. Decreasing the time step
to ensure $k_{max}v_{the}\Delta t<1$ improves the simulation results
(CN with $\Delta t=0.25$), as expected, albeit still with significant
errors. Particle sub-stepping such that $k_{max}v_{the}\Delta\tau<1$
also improves the simulation results, as is demonstrated by the sub-stepping
CN run, but it is still not satisfactory. The implicit solve with
ACC mover is the only one that is able to accurately capture the dynamics
despite the long $\Delta t$ employed, as will be shown later in this
section.

Figure (\ref{fig:IAW-3-movers})b shows the evolution of the electron
kinetic energy. The relative magnitude of the perturbation in the
electron kinetic energy is much smaller than that in the ion kinetic
energy, suggesting that small errors in the electron kinetics are
sufficient to ruin the accuracy of the ion wave.

Figure (\ref{fig:IAW-3-movers})c and (\ref{fig:IAW-3-movers})d show
the time evolution of the conserved quantities of the system, namely,
total energy and total momentum, respectively. Clearly, energy is
conserved exactly, while momentum is not. Similar to the two-stream
instability case {[}Fig.(\ref{fig:compare-2stream}){]}, we see significant
improvement in the momentum conservation when exact charge conservation
is enforced by the ACC mover. With other movers, momentum errors accumulate
in time, and the system can develop nonphysical drifts, which eventually
compromise the accuracy of the IAW dynamics. 

\begin{figure}
\includegraphics[scale=0.8]{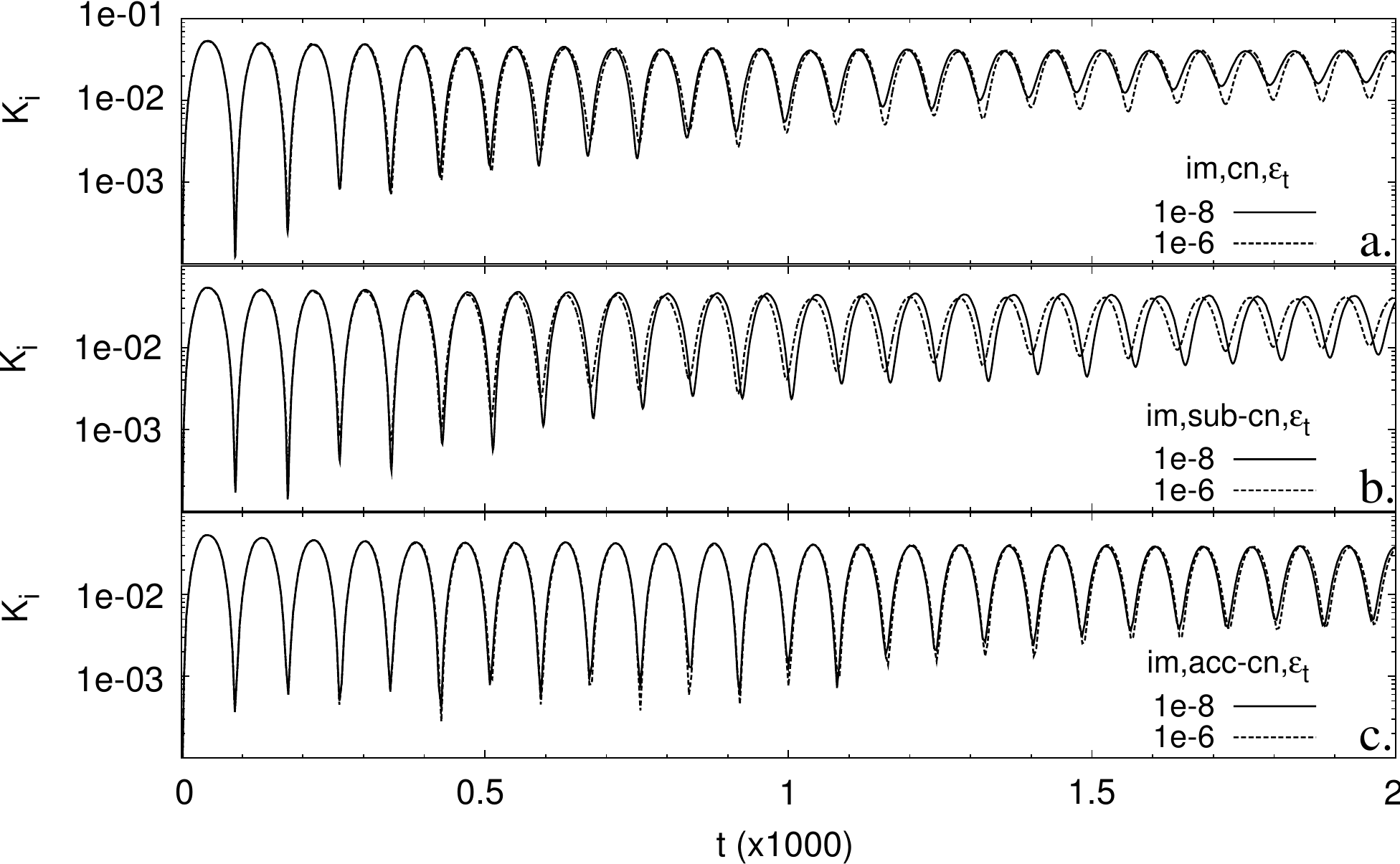}

\caption{\label{fig:Non-tol-IAW}Comparison of the ion kinetic energy history
using two nonlinear tolerances ($10^{-8}$ and $10^{-6}$) in the
implicit simulations. Three particle movers are tested: simple CN
mover (cn), sub-stepping CN mover (sub-CN) and adaptive-charge-conserving
CN mover (ACC-CN). The cn mover and sub-cn mover show strong sensitivity
to the nonlinear tolerance (a, b sub-panels), illustrating the effect
of accumulated errors in particle orbits. In contrast, the acc-cn
particle mover is mostly insensitive to changes in the nonlinear tolerance. }

\end{figure}

It is instructive at this point to compare the implicit ACC-CN solver
with the explicit VP solver. For the latter, we have employed explicit
time steps of $\Delta t=0.1$ and $\Delta t=0.005$, both of which
resolve well the plasma frequency (i.e., the CFL condition is satisfied).
The results of the comparison are shown in Fig.(\ref{fig:IAW-ex-im}).
Most strikingly, only the explicit simulation with the smaller time
step agrees well with the implicit ACC solver, despite the fact that
the latter takes a very large time step for the field equations ($\Delta t=4$,
or $\omega\Delta t\simeq0.14$, $800$ times larger than the explicit
one). This suggests that $v_{th}$-based CFL limits, while sufficient
for stability, may not be enough to guarantee accuracy. In fact, a
$v_{th}$-based CFL can introduce large orbit integration errors for
fast electrons, which may accumulate in time. The only way to control
these long-term errors is to further reduce the explicit time step
to resolve the fast electron orbits, thus exacerbating the computational
efficiency issues of explicit PIC.

The robustness of the implicit solver with respect to the nonlinear
tolerance $\varepsilon_{t}$ for the various movers is examined next.
This is of interest because the level of energy conservation depends
on $\varepsilon_{t}$. Figure (\ref{fig:Non-tol-IAW}) shows the comparison
of simulations using two nonlinear tolerances: $\epsilon_{t}=10^{-8}$
and $\epsilon_{t}=10^{-6}$. Even though the energy conservation is
not exact with $\varepsilon_{t}=10^{-6}$, the total energy is still
conserved well (with errors per time step $\sim10^{-7}$). We find
that the ACC mover is the most robust with respect to changes in the
nonlinear tolerance. This again stresses the point that orbit accuracy,
charge-conservation, and the avoidance of particle tunneling are key
to the overall long-term accuracy of the implicit solver.

\subsection{CPU gain of the implicit scheme vs. the explicit scheme}

Having demonstrated the accuracy of the present implicit scheme, we
proceed to argue and demonstrate by numerical experiment that the
implicit scheme can also be significantly more efficient than the
explicit one, even without a preconditioner.

As has been demonstrated, the proposed energy- and charge-conserving
implicit PIC scheme is not limited to the stringent stability constrains
featured by conventional explicit PIC schemes, namely, $\omega_{pe}\Delta t<1$
and $\Delta x<\lambda_{D}$. Instead, we can choose the time step
to resolve the dynamical timescale of interest, which is in many cases
much slower than the plasma frequency. Similarly, we can choose the
grid size to resolve the relevant spatial scales, which can be much
larger than the Debye length. Thus, there is significant potential
for CPU speed-up of implicit vs. explicit PIC schemes.

We have designed the particle integration scheme to follow orbits
accurately. At first glance, it may appear that the cost of orbit
sub-stepping would offset any potential efficiency gains of the implicit
approach. This would certainly be the case if electron sub-time steps
were comparable to their explicit CFL. However, sub-time steps in
our implicit scheme are not limited by the CFL condition. Instead,
sub-steps are typically $\Delta\tau\lesssim\Delta x/v_{p}$, which
can still be much larger than the explicit CFL when $\Delta x\gg\lambda_{D}$,
and only become restrictive for very few fast particles. It follows
that the implicit scheme should still able to deliver large CPU gains
with respect to explicit, despite particle sub-stepping. 

A back-of-the-envelope estimate confirms that this is the case when
$k\lambda_{D}\ll1$ (the relevant physics regime in many applications
of interest). We begin to compute the expected CPU gain by estimating
the CPU cost for a given PIC solver as:\begin{equation}
CPU=\frac{\Delta T}{\Delta t}n_{p}\left(\frac{L}{\Delta x}\right)^{d}C,\label{eq:CPU-estimation}\end{equation}
where $\Delta T$ is the time-span of the simulation, $n_{p}$ is
the number of particles per cell, ($L/\Delta x$) is the number of
cells, $d$ is the number of physical dimensions, and $C$ is the
computational complexity of the solver employed, measured in units
of a standard explicit PIC VP leap-frogged time step. The implicit-to-explicit
speedup is thus given by:\[
\frac{CPU_{ex}}{CPU_{im}}\sim\left(\frac{\Delta x_{im}}{\Delta x_{ex}}\right)^{d}\left(\frac{\Delta t_{im}}{\Delta t_{ex}}\right)\frac{1}{C_{im}}.\]
For simplicity, we assume that all particles take a fixed sub-timestep
$\Delta\tau$ in the implicit scheme, and that the cost of one time
step with the explicit PIC solver is comparable to that of a single
implicit sub-step (a good approximation when both are dominated by
the cost of moving particles). It follows that $C_{im}\sim N_{FE}\left(\Delta t/\Delta\tau\right)_{im}$,
i.e., the cost of the implicit solver exceeds that of the explicit
solver by the number of function evaluations per time step $N_{FE}$
multiplied by the number of particle sub-steps $\left(\Delta t/\Delta\tau\right)_{im}$.
The number of function evaluations per time step $N_{FE}$ is roughly
equal to the sum of the number of Krylov iterations plus the number
of Newton iterations in a given time step. If an optimal preconditioner
is available, $N_{FE}$ can be made independent of problem dimension
and size (see e.g. \citep{chacon-JCP-rmhd,chacon-JCP-hall,chacon-pop-08-3dmhd}
for examples in the context of fluid modeling of plasmas).

Assuming typical values for $\Delta\tau_{im}\sim0.1\Delta x/v_{th}$,
$\Delta t_{ex}\sim0.1/\omega_{pe}$, $\Delta x_{im}\sim0.2/k$, and
$\Delta x_{ex}\sim\lambda_{D}$, we find that the CPU speedup scales
as:\begin{equation}
\frac{CPU_{ex}}{CPU_{im}}\sim\frac{1}{(k\lambda_{D})^{d+1}}\frac{1}{N_{FE}},\label{eq:CPU-ex-im}\end{equation}
which predicts that large gains are possible when $k\lambda_{D}\ll1$,
particularly in multidimensional applications, but only if $N_{FE}$
is bounded. The latter point underscores the importance of developing
suitable preconditioning strategies. This will be the subject of future
work.

We proceed to verify the scaling of Eq.(\ref{eq:CPU-ex-im}) with
the IAW example. We employ various values of $k\lambda_{D}$ with
both implicit and explicit PIC. Performance results are summarized
in Table \ref{tab:Test-of-CPU}. In all these runs, we have fixed
the number of cells in the implicit runs to 32, and we have used a
constant number of particles per cell (4000). Explicit time steps
and resolutions are chosen for stability, not accuracy. Since the
latter are much more restrictive (as shown earlier in this study),
the results in Table \ref{tab:Test-of-CPU} should be considered as
a conservative lower bound in the efficiency gain potential. As we
decrease $k\lambda_{D}$ (i.e., we increase the domain size), we confirm
that the implicit-to-explicit CPU ratio $CPU_{ex}/CPU_{im}$ (last
column of Table \ref{tab:Test-of-CPU}) increases linearly with the
domain size. This is consistent with the prediction of Eq.(\ref{eq:CPU-ex-im})
when one realizes that the number of function evaluation $N_{FE}$
also increases linearly with the domain size. This results from our
unpreconditioned JFNK implementation combined with the fact that we
have also increased the implicit time step as the domain size grows.
Nevertheless, Table \ref{tab:Test-of-CPU} shows that moderate CPU
gains (up to a factor of 15 for $k\lambda_{D}=0.02$) can be achieved
even without a preconditioner, exemplifying the potential of the implicit
PIC approach for efficient, kinetic plasma simulation.

\begin{table}
\centering{}\caption{\label{tab:Test-of-CPU}Efficiency study of explicit vs. implicit
PIC using the IAW example. We have fixed the number of particles per
cell to 4000, and the resolution of the implicit runs to 32 mesh points. }
\begin{tabular}{cccccc}
$L$ & $k\lambda_{D}$ & $\frac{N_{x}^{ex}}{N_{x}^{im}}$ & $\frac{\Delta t_{im}}{\Delta t_{ex}}$ & $N_{FE}$ & $\frac{CPU_{ex}}{CPU_{im}}$\tabularnewline
\hline
\hline 
10 & 0.628 & 1 & 50 & 13.7 & 0.25\tabularnewline
20 & 0.314 & 2 & 100 & 20 & 0.58\tabularnewline
40 & 0.157 & 4 & 200 & 31.2 & 0.95\tabularnewline
80 & 0.078 & 8 & 200 & 35.8 & 2.18\tabularnewline
160 & 0.039 & 16 & 200 & 43.6 & 5.41\tabularnewline
160 & 0.039 & 16 & 400 & 72.1 & 3.64\tabularnewline
320 & 0.02 & 32 & 200 & 49.6 & 15.4\tabularnewline
320 & 0.02 & 32 & 400 & 67.6 & 11.96\tabularnewline
\end{tabular}
\end{table}

\section{Conclusions}

\label{sec:Conclusion}

In this study, we have undertaken the task of implementing and demonstrating
the accuracy and algorithmic properties of a fully implicit particle-in-cell
solver for plasma simulation. We have focused on a 1D Vlasov-Ampère
electrostatic model as a proof of principle, for which we have developed
an exactly energy-conserving, orbit-averaged discrete formulation.
Exact charge conservation is achieved by an adaptive, charge-conserving,
sub-stepping particle mover. 

The nonlinear residual is converged using an unpreconditioned Jacobian-free
Newton-Krylov solver. A tight nonlinear tolerance is required to achieve
exact energy conservation. Crucial to our implementation is the concept
of particle enslavement, which segregates the particle orbit integration
procedure in the evaluation of the nonlinear residual, and thus allows
a completely general treatment of the particle orbits. This flexibility
has been exploited in this study to implement an adaptive charge-conserving
particle mover, which is compatible with the exact discrete energy
conservation theorem. Since particle quantities are not dependent
variables, the JFNK solver features very modest memory requirements,
comparable to those of fluid simulations (albeit one still needs to
store old-time and new-time particle positions and velocities as auxiliary
variables for the residual computation). Furthermore, being a stand-alone
process, the particle orbit integration stage opens the opportunity
for significant acceleration using dedicated hardware such as GPGPUs.

Numerical experiments with a multiscale problem (the ion acoustic
wave) have demonstrated the superior accuracy properties of the approach.
In particular, we have demonstrated that large implicit time steps
can be used without accuracy degradation, and that explicit approaches
need to use time steps much smaller than the CFL to provide comparable
accuracy. Efficiency-wise, we have argued for the potential to achieve
large CPU gains of implicit methods vs. explicit ones, particularly
in multidimensional applications. We have demonstrated moderate gains
(up to 15) with our unpreconditioned Newton-Krylov implementation
in 1D.

Albeit admittedly at a proof-of-principle level, this study has resolved
a number of limitations of earlier implicit PIC studies, and has uncovered
the potential of fully implicit methods for both \emph{accurate and
efficient} kinetic plasma simulation. Future work will focus on extending
our approach to a fully electromagnetic implementation, on exploring
fully coupled moment-PIC representations that are better suited for
the development of preconditioning strategies, and on exploiting heterogeneous
architectures (e.g., GPGPUs) for the integration of particle orbits.

\section*{Acknowledgments}

The authors would like to acknowledge useful conversations with D.
del-Castillo-Negrete and R. Sánchez. L. C. would like to acknowledge
useful conversations with D. A. Knoll, J. N. Shadid, and J. U. Brackbill,
and the early contributions by G. Lapenta and H. J. Kim while Kim
was a summer student at Los Alamos National Laboratory. This work
has been funded by the Oak Ridge National Laboratory (ORNL) Directed
Research and Development program (LDRD). ORNL is operated by UT-Battelle
for the US Department of Energy under contract DE-AC05-00OR22725.

\newpage{}

\part*{Appendix}

\appendix

\section{Derivation of exact charge-conserving 1D particle mover}

\label{sec:charge-conserv-1d}

We proceed to provide a simple derivation of the 1D charge-conserving
scheme employed in this work, which differs from others reported in
the literature as explained in the main text. As in the main text,
we define the current and charge density $j$ and $\rho$ as\begin{eqnarray}
j_{i} & = & \sum_{p}q_{p}v_{p}S_{m-1}(x_{p}-x_{i})/\Delta x,\label{eq:j}\\
\rho_{i+1/2} & = & \sum_{p}q_{p}S_{m}(x_{p}-x_{i+1/2})/\Delta x,\label{eq:rho}\end{eqnarray}
where $q_{p}$ is the particle charge, $v_{p}$ is the particle velocity,
$S_{m}$ is shape function which is a B-spline of order $m$. Note
that the shape function used for the current density is one order
lower than that for the charge density. In the limit of $\Delta t\rightarrow0,$
it is straightforward to show that \begin{eqnarray*}
\frac{\partial\rho_{i+1/2}}{\partial t} & = & \sum_{p}\frac{q_{p}}{\Delta x}\frac{\partial S_{m}(x_{p}-x_{i+1/2})}{\partial t}\\
 & = & \sum_{p}\frac{q_{p}}{\Delta x}\frac{\partial S_{m}}{\partial x_{p}}\frac{\partial x_{p}}{\partial t}\\
 & = & \sum_{p}\frac{q_{p}}{\Delta x}\frac{S_{m-1}(x_{p}-x_{i})-S_{m-1}(x_{p}-x_{i+1})}{\Delta x}v_{p}\\
 & = & -\frac{j_{i+1}-j_{i}}{\Delta x},\end{eqnarray*}
where $\partial_{x_{p}}S_{m}(x_{p}-x_{i+1/2})=-\partial_{x_{i}}S_{m}(x_{p}-x_{i+1/2})=\left(S_{m-1}(x_{p}-x_{i})-S_{m-1}(x_{p}-x_{i+1})\right)/\Delta x$
is used. With a Crank-Nicolson discretization in time we find that:
\begin{eqnarray*}
\frac{\rho_{i}^{n+1}-\rho_{i}^{n}}{\Delta t} & = & \sum_{p}\frac{q_{p}}{\Delta x}\frac{S_{m}(x_{p}^{n+1}-x_{i})-S_{m}(x_{p}^{n}-x_{i})}{\Delta t}\\
 & = & \sum_{p}\frac{q_{p}}{\Delta x}\frac{\frac{\partial S_{m}(x_{p}^{n+1/2}-x_{i})}{\partial x_{p}}\left(x_{p}^{n+1}-x_{p}^{n}\right)}{\Delta t}\\
 & = & \sum_{p}\frac{q_{p}}{\Delta x}v_{p}^{n+1/2}\frac{S_{m-1}(x_{p}^{n+1/2}-x_{i})-S_{m-1}(x_{p}^{n+1/2}-x_{i+1})}{\Delta x}\\
 & = & -\frac{j_{i+1}^{n+1/2}-j_{i}^{n+1/2}}{\Delta x},\end{eqnarray*}
where we have Taylor expanded $S_{m}(x_{p}^{n+1}-x_{i})$ and $S_{m}(x_{p}^{n}-x_{i})$
about $x_{p}^{n+1/2}$ . 

The exact charge conservation is valid only when the particle is moving
within a cell and $m\leq2$, because the 2nd-order derivative terms
cancel exactly, and no higher-order terms are present in the expansions.
This is seen as follows. \begin{eqnarray}
S_{m}(x_{p}^{n+1}-x_{i}) & = & S_{m}(x_{p}^{n+1/2}-x_{i})+\frac{\partial S_{m}}{\partial x_{p}}(x_{p}^{n+1}-x_{p}^{n+1/2})+\nonumber \\
 &  & +\frac{\partial^{2}S_{m}}{\partial x_{p}^{2}}\frac{(x_{p}^{n+1}-x_{p}^{n+1/2})^{2}}{2}\label{eq:expand-xnp1}\\
S_{m}(x_{p}^{n}-x_{i}) & = & S_{m}(x_{p}^{n+1/2}-x_{i})+\frac{\partial S_{m}}{\partial x_{p}}(x_{p}^{n}-x_{p}^{n+1/2})+\nonumber \\
 &  & +\frac{\partial^{2}S_{m}}{\partial x_{p}^{2}}\frac{(x_{p}^{n}-x_{p}^{n+1/2})^{2}}{2}\label{eq:expand-xpn}\end{eqnarray}
Subtracting Eq.(\ref{eq:expand-xpn}) from Eq.(\ref{eq:expand-xnp1}),
we find exactly that:\[
S_{m}(x_{p}^{n+1}-x_{i})-S_{m}(x_{p}^{n}-x_{i})=\frac{\partial S_{m}}{\partial x}(x_{p}^{n+1}-x_{p}^{n}),\]
which has been used in the derivation above. Since the second derivative
of the B-spline is piecewise continuous only within a cell, it is
required that particles stop at cell boundaries.

\section{Local error estimate of particle orbit integrator}

\label{sec:le-estimate}

We proceed to outline the derivation of the local error estimate for
the particle orbit equations, required for Eq.(\ref{eq:localerror})
in the main text. In what follows, we omit the subscript {}``p''
in particle quantities for convenience. Let us consider the initial
value problem for the particle orbits:\begin{eqnarray*}
\frac{\partial x}{\partial t} & = & v(t,x),\: x(t=0)=x^{0},\\
\frac{\partial v}{\partial t} & = & a(t,x),\, v(t=0)=v^{0}.\end{eqnarray*}
A forward Euler's temporal discretization method gives:\begin{eqnarray}
x_{E}^{\nu+1} & = & x^{\nu}+v(t^{\nu},x^{\nu})\Delta\tau\label{eq:fe-position}\\
v_{E}^{\nu+1} & = & v^{\nu}+a(t^{\nu},x^{\nu})\Delta\tau.\label{eq:fe-velocity}\end{eqnarray}
Using the forward Euler's method as the predictor in a second-order,
time-centered discretization gives:\begin{eqnarray*}
x_{H}^{\nu+1} & = & x^{\nu}+\frac{\Delta\tau}{2}[v(t^{\nu},x^{\nu})+v(t^{\nu+1},x_{E}^{\nu+1})],\\
v_{H}^{\nu+1} & = & v^{\nu}+\frac{\Delta\tau}{2}[a(t^{\nu},x^{\nu})+a(t^{\nu+1},x_{E}^{\nu+1})],\end{eqnarray*}
which is the so-called Heun's formula. A Taylor expansion readily
shows that forward Euler is first-order accurate, while Heun's formula
is second-order accurate:\begin{eqnarray}
x_{E}^{\nu+1} & = & x^{\nu}+\Delta\tau v(t^{\nu},x^{\nu})+O(\Delta\tau^{2}),\label{eq:Taylor-Euler}\\
x_{H}^{\nu+1} & = & x^{\nu}+\frac{\Delta\tau}{2}[v(t^{\nu},x^{\nu})+v(t^{\nu+1},x_{E}^{\nu+1})]+O(\Delta\tau^{3}),\label{eq:Taylor-Heun}\end{eqnarray}
The local error introduced in the position by the Euler step can be
readily estimated by subtracting these two expansions, to find \citep{shampine1977stiffness}:\begin{equation}
le_{x}=[v(t^{\nu+1},x_{E}^{\nu+1})-v(t^{\nu},x^{\nu})]\Delta\tau/2+O(\Delta\tau^{3})\approx\frac{\Delta\tau^{2}}{2}a(t^{\nu},x^{\nu}),\label{eq:localerror-x}\end{equation}
where we have used Eq.(\ref{eq:fe-velocity}) for the last step. Similarly,
we find the following error estimate for the velocity:\begin{equation}
le_{v}=[a(t^{\nu+1},x_{E}^{\nu+1})-a(t^{\nu},x^{\nu})]\Delta\tau/2+O(\Delta\tau^{3})\approx\frac{\Delta\tau^{2}}{2}\left(\frac{\partial a}{\partial x}v\right)^{\nu},\label{eq:localerror-v}\end{equation}
where we have Taylor-expanded the acceleration in the last step. Equations
(\ref{eq:localerror-x}) and (\ref{eq:localerror-v}) are reported
in the main text.

\bibliographystyle{elsarticle-num}
\bibliography{kinetic}

\end{document}